\documentclass[aps,prc,reprint,nofootinbib,superscriptaddress,reprint]{revtex4-2}
\usepackage{graphicx}
\usepackage{dcolumn}
\usepackage{bm}
\usepackage{amssymb}
\usepackage{amsmath}
\usepackage{amsfonts}
\usepackage{xcolor}
\usepackage{multirow}
\usepackage{float}
\usepackage{tabularx}



\def\empile#1\over#2{\mathrel{\mathop{\kern 0pt#1}\limits_{#2}}}

\def\beq{\begin{equation}}
\def\eeq{\end{equation}}
\def\bea{\begin{eqnarray}}
\def\eea{\end{eqnarray}}
\definecolor{darkblue}{rgb}{0.0, 0.0, 0.55}
\definecolor{darkcandyapplered}{rgb}{0.64, 0.0, 0.0}
\def\d3p{\frac{d^3\p}{(2\pi)^3}E_\p}

\newcommand{\Lb}{\left(}
\newcommand{\Rb}{\right)}

\begin{document}


\title{\Large Advanced extraction of the deuteron charge radius from \\
electron-deuteron scattering data}



\author{Jingyi Zhou}
\affiliation{Department of Physics, Duke University, Durham, NC 27708, USA}
\affiliation{Triangle Universities Nuclear Laboratory, Durham, NC 27708, USA}

\author{Vladimir Khachatryan}
\email{vladimir.khachatryan@duke.edu}
\affiliation{Department of Physics, Duke University, Durham, NC 27708, USA}
\affiliation{Triangle Universities Nuclear Laboratory, Durham, NC 27708, USA}

\author{Haiyan Gao}
\affiliation{Department of Physics, Duke University, Durham, NC 27708, USA}
\affiliation{Triangle Universities Nuclear Laboratory, Durham, NC 27708, USA}

\author{Douglas W. Higinbotham}
\affiliation{Thomas Jefferson National Accelerator Facility, Newport News, VA 23606, USA}

\author{Asia Parker}
\affiliation{Department of Physics, Duquesne University, Pittsburgh, PA 15282, USA}

\author{Xinzhan Bai}
\affiliation{Department of Physics, University of Virginia, Charlottesville, VA 22904, USA}

\author{Dipangkar Dutta}
\affiliation{Department of Physics and Astronomy, Mississippi State University, Starkville, MS 39762, USA}

\author{Ashot Gasparian}
\affiliation{Department of Physics, North Carolina A\,$\&$T State University, Greensboro, NC 27411, USA}

\author{Kondo Gnanvo}
\affiliation{Department of Physics, University of Virginia, Charlottesville, VA 22904, USA}

\author{Mahbub Khandaker}
\affiliation{Energy Systems, Davis, CA 95616, USA}

\author{Nilanga Liyanage}
\affiliation{Department of Physics, University of Virginia, Charlottesville, VA 22904, USA}

\author{Eugene Pasyuk}
\affiliation{Thomas Jefferson National Accelerator Facility, Newport News, VA 23606, USA}

\author{Chao Peng}
\affiliation{Physics Division, Argonne National Laboratory, Lemont, IL 60439, USA}

\author{Weizhi Xiong}
\affiliation{Department of Physics, Syracuse University, Syracuse, NY 13244, USA}

\begin{abstract}
\vskip 0.0truecm

To extract the charge radius of the proton, $r_{p}$, from the electron scattering data, the PRad collaboration at Jefferson 
Lab has developed a rigorous framework for finding the best functional forms - the fitters - for a robust extraction of 
$r_{p}$ from a wide variety of sample functions for the range and uncertainties of the PRad data. In this paper we utilize 
and further develop this framework. Herein we discuss methods for searching for the best fitter candidates as well as a 
procedure for testing the robustness of extraction of the deuteron charge radius, $r_{d}$, from parametrizations based on 
elastic electron-deuteron scattering data. The ansatz proposed in this paper for the robust extraction of $r_{d}$, for the 
proposed low-$Q^{2}$ DRad experiment at Jefferson Lab, can be further improved once there are more data. 
\end{abstract}

\maketitle

\section{\label{sec:Intro} Introduction}

Nucleons (protons and neutrons) are the building blocks of atomic nuclei, the structure of which provides an excellent 
laboratory to advance our understanding about how quantum chromodynamics (QCD) -- the theory of strong interaction --  works 
in the nonperturbative region quantitatively where currently our knowledge is rather poor. The proton root-mean-square (rms) 
charge radius -- defined as
\beq
r_{p} \equiv r_{p,rms} \equiv \sqrt{\langle r^{2} \rangle} = \left( -6  \left. \frac{\mathrm{d} G_{E}^{p}(Q^2)}
{\mathrm{d}Q^2} \right|_{Q^{2}=0} \right)^{1/2} ,
\label{eq:eqn_rp}
\eeq
with $G_{E}^{p}$ being the proton electric form factor and $Q^{2}$ the four-momentum transfer squared measured in lepton scattering 
experiments -- also has a major impact on bound-state quantum electrodynamics calculations of atomic energy levels. 
As such the proton charge radius defined in the same way as in lepton scattering experiments~\cite{Miller:2018ybm} can be 
determined from hydrogen spectroscopic measurements. However, there are distinct discrepancies in the measurement results,
observed among three types of experiments. The discrepancies mostly arose after 2010, when high-precision muonic 
hydrogen ($\mu$H) spectroscopy experiments reported two values of $r_{p}$, being $0.8418 \pm 0.0007$~fm 
\cite{Pohl:2010} and $0.8409 \pm 0.0004$~fm \cite{Antognini:1900n}. On the other hand, the world-average value from CODATA-2014
-- $r_{p} = 0.8751 \pm 0.0061$~fm \cite{Mohr:2015ccw} -- determined from atomic hydrogen ($e$H) spectroscopy experiments, 
and the results from electron-proton ($e$-$p$) scattering experiments until 2010 mostly agreed with each other. 
The challenge stemming from such a difference between the $r_{p}$ values, measured from different 
types of the experiments, is known as the {\it proton charge radius puzzle} \cite{Pohl:2013yb,Carlson:2015jba,Hill:2017wzi}.

In the last few years, four more $r_{p}$ measurements from $e$H spectroscopy have been reported. Within experimental 
uncertainties, the one from \cite{Fleurbaey:2018} is consistent with the previous $e$H spectroscopy results, while the other 
two from \cite{Beyer:2017,Bezginov:2019} support the $\mu$H spectroscopy results. However, the latest result from 
\cite{Grinin:2020} reported $r_p = 0.8482 \pm 0.0038$ ~fm, which exceeds the $\mu$H results by $\sim 1.9\sigma$.

Such an agreement with the $\mu$H spectroscopy results is also observed from the $r_{p}$ measured by the PRad collaboration 
at Jefferson Lab \cite{Xiong:2019} -- $r_{p} = 0.831 \pm 0.007_{\rm stat} \pm 0.012_{\rm syst}$~fm -- that used a magnetic-spectrometer-free, 
calorimeter-based method in an unpolarized elastic $e$-$p$ scattering experiment at very low $Q^{2}$, down to 
$ 2.1\!\times\!10^{-4}$~GeV$^{2}$/c$^{2}$ \cite{Gasparian:2014rna,Peng:2015szv}.

The situation becomes similarly interesting and challenging if we move on to discuss measurements of the rms charge 
radius of the deuteron, $r_{d}$, in electron-deuteron ($e$-$d$) scattering experiments as well as in $eD$ and $\mu D$ spectroscopy. 
In particular, the CREMA collaboration has reported a deuteron charge radius -- $r_{d} = 2.12562 \pm 0.00078$~fm -- from a 
muonic spectroscopy-based measurement of three $2P \rightarrow 2S$ transitions in $\mu D$ atoms \cite{Pohl:2016}, which 
is 2.7 times more accurate but 7.5-$\sigma$ smaller than the CODATA-2010 world-average value \cite{Mohr:2012}. The radius from \cite{Pohl:2016} is 
also 3.5-$\sigma$ smaller than the $r_{d}$ value, $2.1415 \pm 0.0045$~fm, extracted from an electronic 
spectroscopy-based measurement \cite{Pohl:2016glp} of $1S \rightarrow 2S$ transitions in $eD$ atoms, after these 
transitions have already been measured in \cite{Parthey:2010aya}.

Thereby, one also observes discrepancies from $r_{d}$ measurements (like in the case of $r_{p}$) that have given rise to 
another challenge, dubbed as the {\it deuteron charge radius puzzle}. The PRad collaboration has proposed a low-$Q^{2}$ 
unpolarized elastic $e$-$d$ scattering experiment named as DRad -- basically anchored upon PRad's 
experimental setup -- for a model-independent extraction of $r_{d}$ with a subpercent $(\leq 0.25\%)$ precision, in order 
to address this newly developed puzzle \cite{DRad}. 

Thus, given the importance of measuring not only $r_{p}$ but also $r_{d}$, our goal is to show how one can robustly extract 
$r_{d}$ and control its uncertainties in a fitting procedure, using four parametrizations of the deuteron charge form factor, 
 $G_{C}^{d}$ \cite{Abbott:2000ak,Abbott:2000fg,kobushkin1995deuteron,Parker:2020,Sick:1974suq,Zhou:2020}.
In this paper we apply and extend the ansatz used in \cite{Yan:2018bez}, in which a comprehensive and
systematic method is presented for choosing mathematical functions that can robustly extract $r_{p}$ from a broad set of 
input functions describing the proton electric form factor, $G_{E}^{p}$.  

The rest of the paper is presented as follows. Sec.~\ref{sec:FormFacRad} has a brief discussion on the 
deuteron form factors and the radius extraction. In Sec.~\ref{sec:Fitting} we describe the general fitting procedure on how 
to extract $r_{d}$ from generated $G_{C}^{d}$ pseudo-data in the DRad kinematics and define some quantities to compare the 
properties of different fitters. In Sec.~\ref{sec:RobFitCan} we introduce the pseudo-data generation from the $G_{C}^{d}$ 
parametrizations and discuss the method for searching for a fitter that will be able to extract $r_{d}$ by using the available 
elastic $e$-$d$ scattering data. In Sec.~\ref{sec:Robustness} we show a comprehensive way to estimate the bias for 
$r_{d}$ extraction. We conclude on our paper and discuss its prospects at the end. Also, in the Appendices we discuss the results of 
testing a few theoretical models and provide another robust fitter candidate which is analogous to the one considered in 
Sec.~\ref{sec:RobFitCan}.

\section{\label{sec:FormFacRad} Form factors and charge radius from unpolarized elastic electron-deuteron cross section}

The understanding of the electromagnetic properties of the deuteron is of fundamental importance in nuclear physics, 
given that the deuteron is the only bound two-nucleon system. It is expected that at the low-$Q^{2}$ region, where the 
relativistic effects and non-nucleonic degrees of freedom are expected to be negligible, 
the deuteron form factors are dominated by part of its wave function for which the two constituent nucleons are far apart. 
Theoretical calculations of $r_{d}$ are considered to be reliable since they are independent of the nucleon-nucleon potential 
(for a broad class of potentials), and depend mostly on the binding energy and neutron-proton scattering length \cite{Wong:1994sy}. 
This makes $r_{d}$ a perfect observable for a theory-experiment comparison. 

So far three experiments have been conducted 
for determination of $r_{d}$ from unpolarized elastic $e$-$d$ scattering at low $Q^{2}$ \cite{Berard:1974ev,Simon:1981br,Platchkov:1989ch}, 
the cross section of which in the one-photon exchange approximation is given by
\beq
\frac{\mathrm{d}\sigma}{\mathrm{d}\Omega}\Lb E, \theta \Rb  = \sigma_{_{\!NS}} \Lb A_{d}(Q^{2}) + B_{d}(Q^{2})\,
\tan^{2}{\!\Lb \frac{\theta}{2} \Rb} \Rb ,
\label{eq:eqn_sigma}
\eeq
where $\sigma_{_{\!NS}}$ is the differential cross section for the elastic scattering from a pointlike and spinless 
particle at a scattering angle $\theta$ and an incident energy $E$. The four-momentum transfer squared carried 
by the exchanged virtual photon is defined in terms of the four-momenta of the incident ($k$) and scattered 
($k^{\prime}$) electrons: $Q^{2} = -\Lb k - k^{\prime} \Rb^{2}$. In this case the deuteron structure functions 
in Eq.\thinspace(\ref{eq:eqn_sigma}) are related to its charge, $G_{C}^{d}$, magnetic dipole, $G_{M}^{d}$, and 
electric quadrupole, $G_{Q}^{d}$, form factors via \cite{Jankus:1997,Gourdin:1963, Mainz}
\bea
A_{d}(Q^{2}) & = & \Lb G_{C}^{d}(Q^{2}) \Rb^{2} + \frac{2}{3}\,\tau \Lb G_{M}^{d}(Q^{2}) \Rb^{2} +
\nonumber \\
& & + \frac{8}{9}\,\tau^{2} \Lb G_{Q}^{d}(Q^{2}) \Rb^{2} ,
\nonumber \\
B_{d}(Q^{2}) & = & \frac{4}{3}\,\tau (1 + \tau) \Lb G_{M}^{d}(Q^{2}) \Rb^{2} ,
\label{eq:eqn_deuteronstrucfunc}
\eea
with $\tau = Q^{2}/4M_{d}^{2}$, where $M_{d}$ is the deuteron mass. Also, there are the following additional relations:
\begin{displaymath}
G_{C}^{d}(0) = 1 ,\,\,\,\,\,\,\,\,\frac{G_{Q}^{d}(0)}{\mu_{Q}^{d}} = 1 ,\,\,\,\,\,\,\,\frac{G_{M}^{d}(0)}{\mu_{M}^{d}} = 1 ,
\label{eq:eqn_deuteronstrucfunc2}
\end{displaymath}
with the given deuteron electric quadrupole moment, $\mu_{Q}^{d}$, and magnetic dipole moment, 
$\mu_{M}^{d}$\footnote{Throughout the text we use dimensionless $\mu_{M}^{d} \equiv (\mu_{M}^{d}/\mu_{N}$) = 0.8574 
and $\mu_{Q}^{d} \equiv (\mu_{Q}^{d}/{\rm fm^{2}})$ = 0.2859 \cite{Garcon:2001sz}.}.

At very low but experimentally accessible $Q^{2}$ such as $\sim 10^{-4}~({\rm GeV/c})^{2}$, 
the contributions from $G_{Q}^{d}$ and $G_{M}^{d}$ to the scattering process are negligible. By choosing different 
$G_{M}^{d}$ and $G_{Q}^{d}$ form factors  \cite{Abbott:2000ak,Abbott:2000fg,kobushkin1995deuteron,Parker:2020,Sick:1974suq,Zhou:2020} from 
four data-driven models discussed in Appendix A (and throughout the paper) for extracting $G_{C}^{d}$ from the cross section, the 
effects of the choice of the form-factor models on the deuteron radius are found to be $0.03$ and $0.009\%$, respectively. Thereby, in order to extract the deuteron rms 
charge radius from $e$-$d$ scattering data, one should fit $G_{C}^{d}$ to the experimental data as a function of \(Q^2\), and calculate 
the slope of this function at \(Q^2=0\), according to
\beq
r_{d} \equiv r_{d,rms} \equiv \sqrt{\langle r^{2} \rangle} = \left( -6  \left. \frac{\mathrm{d} G_{C}^{d}(Q^2)}
{\mathrm{d}Q^2} \right|_{Q^{2}=0} \right)^{1/2} ,
\label{eq:eqn_rd}
\eeq
in analogy to how $r_{p}$ is obtained.

\section{\label{sec:Fitting} The fitting procedure and robustness}
\subsection{\label{sec:procedure} The general procedure}

Refs.~\cite{Yan:2018bez,Kraus:2014qua, bernauer2014electric} give a general framework with input form-factor functions 
and various fitting functions, for finding functional forms (fitters) that allow for a robust extraction of an input proton radius.  
Analogously, we can find robust fitters to extract $r_{d}$ by testing all combinations of available input functions and fitting functions. 

From a developed routine\footnote{A C++ coded program library has been created for generating, adding fluctuations to, and fitting the pseudo-data 
\cite{Yan:2018bez,Radius_fitting_lib} (see Sec.~\ref{sec:RobFitCan}). The bin-by-bin and overall type fluctuations are assumed to imitate 
the binning and random uncertainties of a given set of real data. For fitting purposes the library uses the MINUIT package of CERN ROOT 
\cite{Brun:1997,James:1975}.} we generate many sets of $G_{C}^{d}$ pseudo-data values with user-defined fluctuations at given $Q^{2}$ bins 
by using some $G_{C}^{d}$ charge form-factor models as input. Then we use various fitting functions to fit the pseudo-data and extrapolate them 
to $Q^2=0$, in order to obtain the $r_d$ values according to Eq.~(\ref{eq:eqn_rd}).



\begin{figure}[hbt!]
\centering
\includegraphics[width=0.425\textwidth]{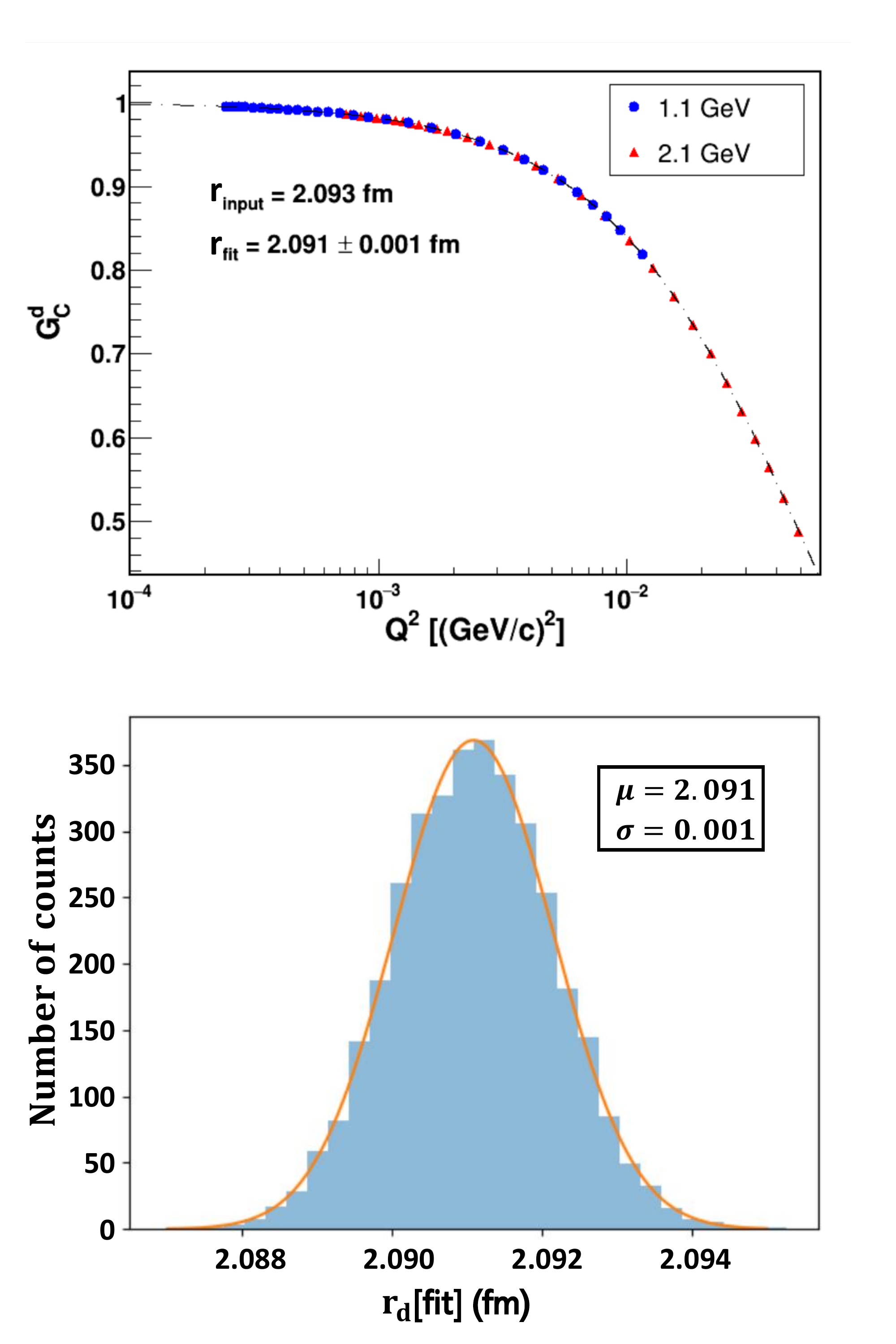}
\caption{(Color online) The upper plot shows an example of one fit using the Abbott1 model (see Sec.~\ref{sec:pseudo-data}) as input and 
Rational\,(1,1) (see Sec.~\ref{sec:RMSE}) as the fitting function. The lower plot shows an example of $r_{d}{\rm [fit]}$ distribution 
obtained by following the above-mentioned pseudo-data and fitting procedure. A Gaussian function, similar to that in Eq.~(\ref{eq:eqn_Gaus}), 
is used to fit the distribution.}
\label{fig:bias_fitter}
\end{figure}

When the program library generates bin-by-bin type fluctuations added to the pseudo-data, it occurs according to the 
user-defined random Gaussian distribution at each bin. Stated otherwise, in order to mimic the bin-by-bin fluctuations 
($Q^{2}$-independent) of the data, the pseudo-data should be smeared by shifting the $G_{C}^{d}$ central 
value at each $Q^{2}$ bin with a random number following the Gaussian distribution, \(\mathcal{N}(\mu , \sigma^{2}_{g})\), given by
\begin{equation}
\mathcal{N}(\mu,\,\sigma_{g}^{2}) = \frac{1}{\sqrt{2\pi\sigma_{g}^{2}}}\,e^{-\left( G_{C}^{d} - \mu \right)^{2}/
\left( 2\sigma_{g}^{2} \right)} .
\label{eq:eqn_Gaus}
\end{equation}
In this paper we take \(\mu = 0\) and \(\sigma_{g} = \delta G_{C}^{d}\), where \(\delta G_{C}^{d}\) comes from the estimated 
statistical and/or systematic uncertainties in the $e$-$d$ (DRad) experiment. The produced tables of $G_{C}^{d}$ vs. $Q^{2}$ with 
fluctuations are fitted with a number of fitters for extracting $r_{d}$ (see Fig.~\ref{fig:bias_fitter}). 

\begin{figure*}[hbt!]
\centering
\includegraphics[scale=0.315]{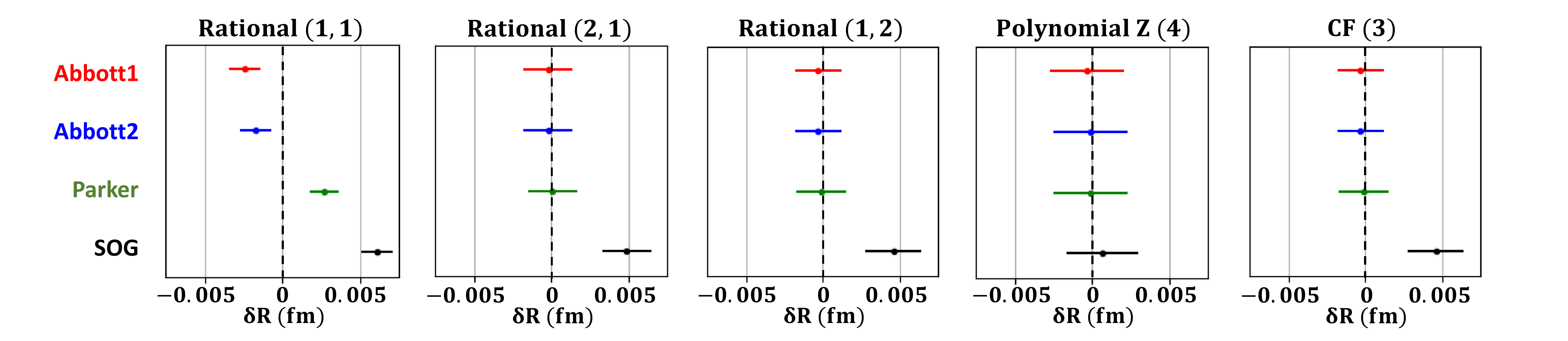}
\caption{(Color online) Five fitters from \cite{Yan:2018bez}, which give the best RMSE values for extraction of $r_{d}$, when they are 
fitted with pseudo-data generated by the four $G_{C}^{d}$ parametrizations that we refer to as Abbott1 and Abbott2 
\cite{Abbott:2000ak,kobushkin1995deuteron}, as well as Parker \cite{Parker:2020} and SOG \cite{Sick:1974suq,Zhou:2020} models. The error bars 
show the statistical uncertainty of the deuteron radius.
}%
\label{fig:DRad_fitter}%
\end{figure*}

\subsection{\label{sec:robustness2} The robustness and goodness of fitters}
In this paper, the robustness of a fit function is determined by its ability to extract 
$r_{d}$ from a variety of pseudo-data generated from plausible form-factor parametrizations. Our conviction is that the true and unknown form-factor function is reasonably approximated by the trial functions. As discussed in \cite{Higinbotham:2019jzd}, descriptive functions 
(such as high-order polynomials), which precisely match onto the data over a limited $Q^2$ range, are often not the same as predictive functions 
(such as low-order rational functions), which are able to extrapolate. Unsurprisingly, the predictive functions are often found to be the most 
robust functions for $r_p$ extractions. 


In order to determine the robustness of a fitter based upon the general procedure already discussed, one can 
compare the size of the bias (${\rm bias}\equiv \delta r_{d}{\rm} = r_{d}[{\rm mean}]$ - $r_{d}[{\rm input}]$) with the variance $\sigma$ 
(the rms value of the radius distribution). The bias
comes from the mismatch of the fitting function and the underlying
generation function, which leads to a misprediction of the slope at
$Q^2=0$. The variance reflects the influence of the $G_{C}^{d}$ bin-by-bin uncertainties on the radius. If $\delta r_{d} < \sigma_{stat}$ 
(statistical variance) for most of the input form-factor models, the given fitter will be considered as sufficiently robust. In the case of an experiment, 
the goal of which is to minimize the overall uncertainty, we should also consider the bias and variance together, using the 
root-mean-square error 
(RMSE) \cite{HTF:2009}\footnote{The RMSE discussed throughout this paper is somewhat different from that discussed in \cite{Yan:2018bez}, where 
the authors have considered $\sigma_{stat}$ in the formula of RMSE.}:

\beq
\textrm{RMSE} = \sqrt{\textrm{$\delta r_{d}$}^{2} + \textrm{$\sigma_{total}$}^{2}} ,
\label{eq:eqn_RMSE}
\eeq
where $\sigma_{total}$ includes both bin-by-bin statistical and systematic uncertainties. The RMSE is a standard way of quantifying goodness of fitters.
The smaller the RMSE is, the better the corresponding fitter is. 
Eventually, we need to find a fitter(s) that can extract the deuteron radius precisely, from pseudo-data generated from a range of plausible form 
factors, which should be reasonable approximations to the unknown true function to allow for the best possible determination of the radius when the 
fitter is applied to $e$-$d$ experimental data. The key point here is that the fitters are determined prior to obtaining the experimental results from the 
planned $Q^2$ range and precision of the DRad experiment.

\subsection{\label{sec:RMSE} Initial studies and motivation}

Ref.~\cite{Yan:2018bez} takes into account different reasonable  approximations to the unknown true function by using nine different $G_{E}^{p}$ 
form-factor parametrizations to generate pseudo-data in the PRad $Q^2$ range. The studies show that the two-parameter rational function, 
Rational\,(1,1), is robust and the best fitter for extraction of $r_{p}$ for the range and uncertainties of the PRad data, represented by
\bea\label{R11}
& & f_{\rm Rational\,(1,1)}(Q^{2}) \equiv {\rm Rational\,(1,1)} = 
\nonumber \\
& &
~~~~~~~~~~~~~~~ = p_{0}\,G_{E}^{p}(Q^{2}) = p_{0} \frac{1 + p_{1}^{(a)}Q^{2}}{1 + p_{1}^{(b)}Q^{2}} ,
\eea
where \(p_{0}\) is a floating normalization parameter, and \(p_{1}^{(a)}\) and \(p_{1}^{(b)}\) are two free fitting parameters. The radius 
is determined by \(r_{p} = \sqrt{6 \left( p_{1}^{(b)} - p_{1}^{(a)} \right)}\). The other two robust fitters are the two-parameter
continued fraction and the second-order polynomial expansion of the so-called Z transformation \cite{Yan:2018bez,Lee:2015jqa}, which can 
extract the input proton radius regardless of the input electric form-factor functions.

Eq.~(\ref{R11}) is actually a special case from the class of the multiparameter rational function of $Q^{2}$ given by
\bea\label{RNM}
& & f_{\rm Rational\,(N,M)}(Q^{2}) \equiv {\rm Rational\,(N,M)} = 
\nonumber \\
& &
~~~~~~~~~~ = p_{0}\,G_{E}^{p}(Q^{2}) = p_{0} \frac{1 + \sum\limits_{i=1}^{N} p_{i}^{(a)}Q^{2i}}
{1 + \sum\limits_{j=1}^{M} p_{j}^{(b)}Q^{2j}} ,
\eea
where the orders $N$ and $M$ are defined by the user.

All the fitters studied in \cite{Yan:2018bez} have been tested here by fitting pseudo-data generated using the four $G_{C}^{d}$ 
parametrizations from  \cite{Abbott:2000ak,kobushkin1995deuteron,Parker:2020,Sick:1974suq,Zhou:2020} (see Sec.~\ref{sec:pseudo-data}). For this test 
we took the DRad kinematic range of $2 \times 10^{-4}~{\rm (GeV/c)^2} < Q^{2} < 0.05~{\rm (GeV/c)^2}$, using bin-by-bin statistical uncertainties 
from $0.02$ to $0.07\%$ and systematic uncertainties from $0.06$ to $0.16\%$. The bias and $\sigma_{stat}$ values of the five best fitters are 
shown in Fig.~\ref{fig:DRad_fitter}\footnote{The three-parameter continued fraction [CF\,(3)] and Polynomial\,Z\,(4) are from the classes 
of the CF expansion and multiparameter polynomial expansion of $Z$, respectively. For their explicit expressions we refer the reader to 
\cite{Yan:2018bez}. The CF\,(3) has the same functional form as Rational\,(1,2).}. Although the four-parameter Polynomial\,Z gives 
the smallest bias, it also gives the largest variance and RMSE amongst them. The RMSE value of Rational\,(1,1) is the smallest one, though it gives 
larger bias compared to the others. 

However, given the limited number of $G_{C}^{d}$ parametrizations, the robustness of the fitters cannot be convincingly determined from these results.
In this case, we can not mimic different kinds of approximations to the unknown true function as comprehensively as it can be done for the proton $G_{E}^{p}$ 
models. We have also studied some theory-based models (discussed in Appendix~B), and found that those models have large 
discrepancies with the experimental data, which show that the testing method of robustness applied to PRad is no longer suitable for the deuteron radius 
extraction. Based on our studies, the bias is a non-negligible source of the $r_{d}$ systematic uncertainty estimated for the DRad experiment. This 
observation was our motivation for looking into other potentially better fitters for DRad, which might give similar variance but smaller bias as compared 
to those of Rational\,(1,1). At the same time, by having limited $G_{C}^{d}$ parametrizations at our disposal, we also need to develop a more comprehensive 
method to estimate the bias when using various fitters.

\section{\label{sec:RobFitCan} Searching for a robust fitter candidate}
\subsection{\label{sec:pseudo-data} Pseudo-data generation}

Here we give some specific details on the pseudo-data generation and fitting procedure described in the previous section: \\

(A) Generating pseudo-data:
\begin{itemize}
\item[(i)]Four $G_C^{d}$ parametrizations based on available experimental data (named as Abbott1, Abbott2, Parker and SOG) 
are used to generate $G_C^{d}$ values at given \(Q^{2}\) bins. The details of these parametrizations are discussed in Appendix A.

\item[(ii)] 30 $G_{C}^{d}$ pseudo-data points at 1.1~GeV and 37 $G_{C}^{d}$ points at 2.2~GeV are generated 
from each of the four deuteron models in step i. Our binning choice for DRad is based on the binning of PRad. There are 30 bins from 0.8$^{\circ}$ to 6.0$^{\circ}$ at 1.1-GeV 
beam energy, and 37 bins from 0.7$^{\circ}$ to 6.0$^{\circ}$ at 2.2~GeV \cite{Binningset}.
\end{itemize}

(B) Adding fluctuations to the pseudo-data and fitting:
The following steps are repeated 10000 times, which is sufficient to obtain stable results of the mean value and
rms of the $r_{d}{\rm [fit]}$ distribution to the precision of $10^{-4}$~fm.
\begin{itemize}
\item[(i)] To add statistical fluctuations, the total 67 pseudo-data points generated in step A are smeared by 
67 different random numbers according to Eq.~(\ref{eq:eqn_Gaus}).

\item[(ii)] In this step a set of pseudo-data is fitted by a specific fitter $f_{d}(Q^{2})$. The data points at 1.1 and 
2.2~GeV are combined and fitted by that fitter with two different floating normalization parameters corresponding to 
these two beam energies. The other fitting parameters in the fitter are required to be the same for the two energies.

\item[(iii)] Then the fitted radius is calculated from the fitted function in step ii, with 
\begin{equation}\label{eq:eqn_rd2}
r_{d}{\rm [fit]} = \left( -6 \left. \frac{\mathrm{d} f_{d}(Q^{2})}{\mathrm{d}Q^{2}} \right|_{Q^{2}= 0} \right)^{1/2}.
\end{equation}

\end{itemize}

\subsection{\label{sec:Datadriven} Data-driven method}

As described in Sec~\ref{sec:RMSE}, our studies have shown that the bias is an important source of systematic uncertainty in 
the extraction of $r_d$. Hence, to better control and/or minimize the bias in the $r_d$ extraction, such as that obtained by the Rational\,(1,1) 
fitter, we propose a data-driven approach to search for a new robust fitter candidate. 

The Rational\,(1,3) is a function with four free parameters that has been used in \cite{kelly2004simple} to fit $G_{E}^{p}$. 
Compared to the Rational\,(1,1), it has 
good asymptotic behaviors satisfying not only $G_{C}^d = 1$ at $Q^2=0$
but also $G_{C}^d \rightarrow 0$ at \(Q^{2} \rightarrow \infty\). This fitter function is given by
\bea\label{R13}
& &
f_{\rm Rational\,(1,3)}(Q^{2}) \equiv {\rm Rational\,(1,3)} = 
\nonumber \\
& & 
~~ = p_{01}\,G_{C}^{d}(Q^{2}) = 
p_{01}\frac{1 + a_{1}Q^{2}}{1 + b_{1}Q^{2} + b_{2}Q^{4} + b_{3}Q^{6}} ,
\eea
where \(a_{1}, b_{1}, b_{2}, b_{3}\) are free parameters, and \(p_{01}\) is a floating normalization parameter. 

In order to control the variance of $r_{d}{\rm [fit]}$, we fit this function to the existing experimental data sets in Table~1 
of \cite{Abbott:2000ak}, which provides $G_{C}^{d}$ and $\delta G_{C}^{d}$ at fixed $Q^2$ values that are typically higher than 
the values of the $Q^2$ range of the proposed DRad experiment. With $\chi^2/{\rm NDF} \simeq 1.25$, we determine \(b_{2} = 0.0416 \pm 0.0152 \) 
and \(b_{3} = 0.00474 \pm 0.000892 \). Then fixing these values for fitting the pseudo-data in the (low-$Q^{2}$) DRad range 
will render a fitter, which we refer to as fixed Rational\,(1,3) or fRational\,(1,3):
\bea\label{fixed_Rational}
& & f_{\rm fixed\,Rational\,(1,3)}(Q^{2}) \equiv {\rm fRational\,(1,3)} =  
\nonumber \\
& & 
~~~~~~ = p_{01}\frac{1 + a_{1}Q^{2}}{1 + b_{1}Q^{2} +  b_{2,{\rm fixed}}Q^{4} + b_{3,{\rm fixed}} Q^{6}} , 
\eea
where the uncertainties in the fixed parameters are taken into account when we calculate the bias.
In principle, if some fitter functions have fitting uncertainties in their fixed parameters, those parameters should be smeared 
using a Gaussian distribution, with $\sigma_{g}$ to be the fitting uncertainty (see Eq.~(\ref{eq:eqn_Gaus})). We repeat this step 
in the fitting procedure 10000 times as discussed in Sec.~\ref{sec:pseudo-data}.

\begin{figure}[hbt!]
\centering
\includegraphics[width=0.425\textwidth]{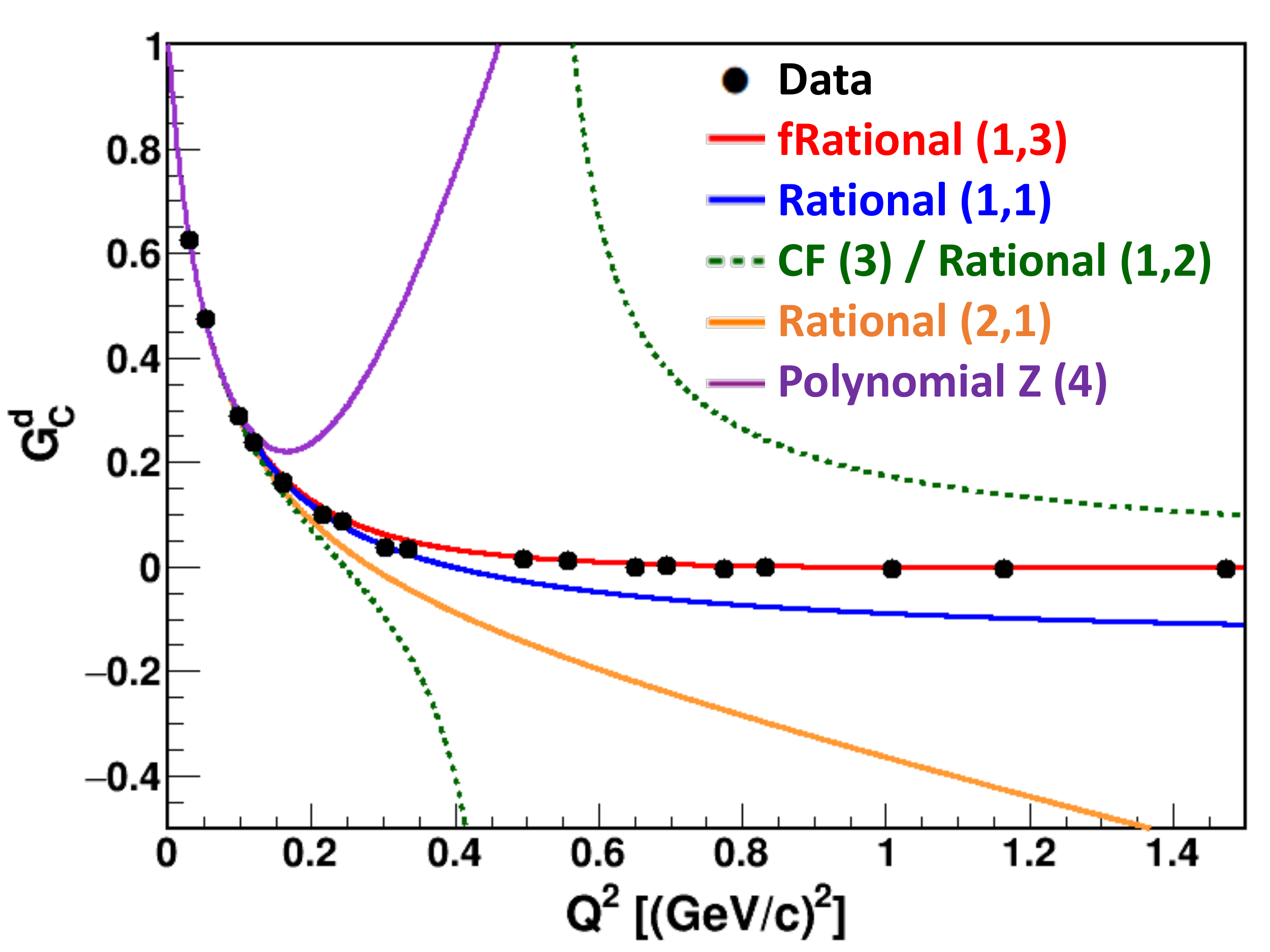}
\includegraphics[width=0.455\textwidth]{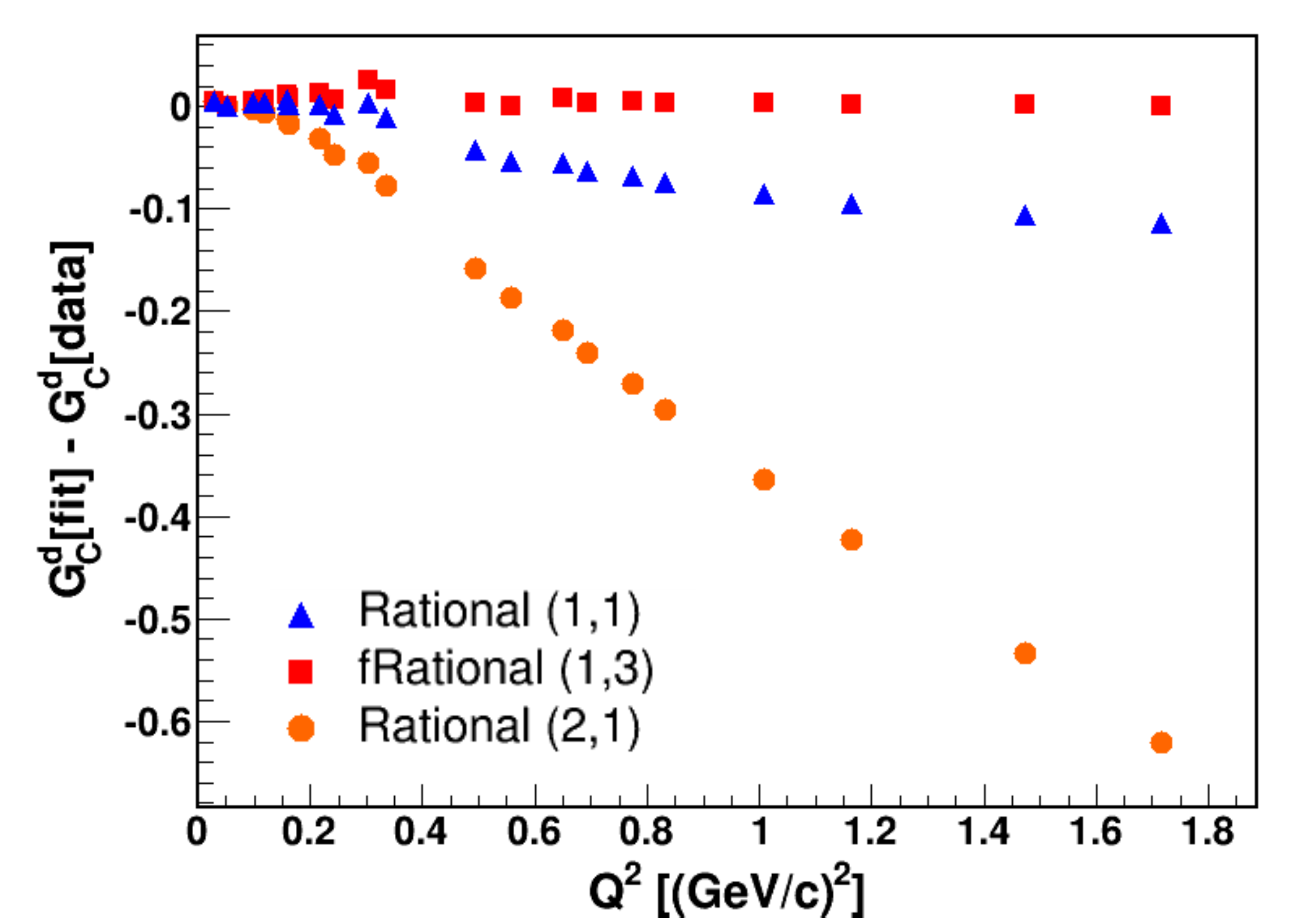}
\caption{(Color online)  The upper plot shows the fRational\,(1,3), Rational\,(1,1), 
Rational\,(1,2), Rational\,(2,1), CF\,(3), and Polynomial\,Z\,(4) obtained from fitting the pseudo-data generated by the Abbott1 model \cite{Abbott:2000ak},
which for comparison are overlaid with the black colored data points listed in Table 1 of \cite{Abbott:2000ak}. The color coding is displayed in the 
legends, where the CF\,(3) and Rational\,(1,2) are the same and described by the two asymptotic green dotted lines. The lower plot shows the residual 
points for the fRational\,(1,3), Rational\,(1,1), and Rational\,(2,1), where ``the residual" means the difference between $G_{C}^{d}[{\rm fit}]$ described 
by the fitters and $G_{C}^{d}[{\rm data}]$ from the data.}

\label{fig:R11_vs_FixedR13}
\end{figure}

To compare the differences between Rational\,(1,1), fRational\,(1,3) and other fitters shown in Fig.~\ref{fig:DRad_fitter}, all the functions 
are plotted in the Abbott1/Abbott2 model range [from $Q^2$ = \(3 \times 10^{-2}\ {\rm to}\ 1.5~{\rm (GeV/c)^2}\)]. 
The parameters in these fitters are determined by fitting pseudo-data generated from the Abbott1 model in the DRad $Q^2$ range. 
The results from the Abbott2, Parker, and SOG models are very similar, therefore we do not show them here. As shown in Fig.~\ref{fig:R11_vs_FixedR13}, 
all the fitters describe the data quite well in the low-\(Q^2\) range [$Q^2 < 0.15~{\rm (GeV/c)^2}$],  
while Polynomial\,Z\,(4) and 
CF\,(3) diverge. At high-\(Q^2\) range, the fRational\,(1,3) describes the data much better than the other fitters, which means that the 
fRational\,(1,3) has a better asymptotic behavior at high \(Q^2\). Based on this observation, the fRational\,(1,3) may also have a potential to 
describe the data in the low-\(Q^2\) range better than the Rational\,(1,1). Other than the fRational\,(1,3) functional form, we 
have also studied another fitter, which has similar properties and is capable of extracting $r_{d}$ robustly. The details on our 
studies for this fitter are presented in Appendix~C.

\section{\label{sec:Robustness} A comprehensive way to estimate the bias in deuteron charge radius extraction}

\subsection{\label{sec:Smearing} Smearing procedure}
After the candidate fitter is found, the robustness for the deuteron radius extraction needs to be tested. Being limited by the number of 
$G_{C}^{d}$ parametrizations, in order to reflect various reasonable approximations to the unknown true function, the 
parameters in the two Abbott as well as in the Parker and SOG models should be smeared. Once they are smeared, the functional forms describing 
the models are different, and are used to perform a variety of extrapolations at low \(Q^2\). Overall, this test is a \(\chi^{2}\) test, 
which consists of the following steps.

(A) Smearing of the parameters and calculation of \(\chi^{2}\):
First, we smear all the parameters for $\pm 10\%$, following a uniform distribution in a model. 
Then we use the smeared model to generate the corresponding \(G_{C}^{d~\prime}\) with respect to its value at the same \(Q^{2}\) bin in the (\(Q^{2}\), 
\(G_{C}^{d}\), \(\delta G_{C}^{d}\)) data set from Table~1 of \cite{Abbott:2000ak}. Afterwards, we calculate \(\chi^{2}\) by
\begin{equation}\label{Eq:chi2}
\chi^{2} = \sum {\frac{(G_{C}^{d} - G_{C}^{d~\prime})^{2}}{\delta G_{C}^{d}}} .
\end{equation}

(B) Checking of the acceptable region:
The definition of an acceptable \(\chi^{2}\) region is that the probability of the calculated \(\chi^{2}\) (after the parameters are 
smeared) with a specific degree of freedom is ``acceptable'' when it is larger than 99.7\% in the $\chi^{2}$ probability distribution. 
This requirement restricts the value of \(\chi^{2}\), which means that the smeared model should not be far away from the real 
experimental data. With the specific degree of freedom \(\nu^{2}\), the \(\chi^{2}\) probability distribution is defined as
\begin{equation}\label{chi2pro}
f(\chi^{2}) = \frac{1}{2^{\nu/2}\Gamma(\nu/2)} e^{(-\chi^{2}/2)}(\chi^{2})^{(\nu/2) - 1} .
\end{equation}
Integrating the function in Eq.~(\ref{chi2pro}), from zero to \(\chi^{2}_{0}\), will result in the probability for \(\chi^{2}_{0}\). 
The number of degrees of freedom (NDF) and the critical $\chi^2_0$ value for each of the four smeared 
data-based models are shown in Table.~\ref{tab:table_chi2first}.
\begin{table}[h!]
  \begin{tabularx}{0.425\textwidth}{X X c}
      \hline
      \hline
      Model & \!\!\!NDF & $\chi^2_0$\\
      \hline
    {Abbott1} & 16 & 35.9 \\
    {Abbott2} & 7  & 21.6\\
    {Parker} &  16 & 35.9\\
    {SOG} &     11 & 28.2\\
      \hline
      \hline
  \end{tabularx}
      \caption{The number of degrees of freedom and the critical $\chi^2_0$ value for each of the four smeared data-based models.}
\label{tab:table_chi2first}
\end{table}

If the calculated \(\chi^{2}\) is smaller than the above numbers for each smeared model, then we keep the given smeared 
model and go to the next step. For each smeared model there is a new $r_{d}{\rm [input]}$, which is calculated by 
Eq.~(\ref{eq:eqn_rd}) with the slope of a smeared model at \(Q^{2} = 0\). If \(\chi^{2}\) is unacceptable, the parameters of the model 
are re-smeared and the whole procedure is repeated.

(C) Generating pseudo-data: If the smeared models pass step B, in this case these models can be utilized to generate 
sets of pseudo-data in the DRad $Q^{2}$ range using the binning discussed in Sec.~\ref{sec:RobFitCan}. 

(D) Fitting and calculating the bias:
After the pseudo-data are generated, we use the selected fitter to fit and obtain the quantity $r_{d}{\rm [fit]}$. 

(E) Repeating and obtaining the relative bias:
In this step the above procedure for each model is repeated 10000 times for obtaining 10000 values of relative bias, which is defined as
$\delta r_{d}/r_{d}{\rm [input]}$.

(F) Finalization:
From each relative bias distribution of the smeared Abbott1, Abbott2, Parker, and SOG models, we select the rms value to 
calculate $\delta r_{d}$ in Eq.~(\ref{eq:eqn_RMSE}).

\subsection{Proof of the robustness test using the proton form-factor models}

\begin{figure*}[hbt!]
\centering
\includegraphics[scale=0.615]{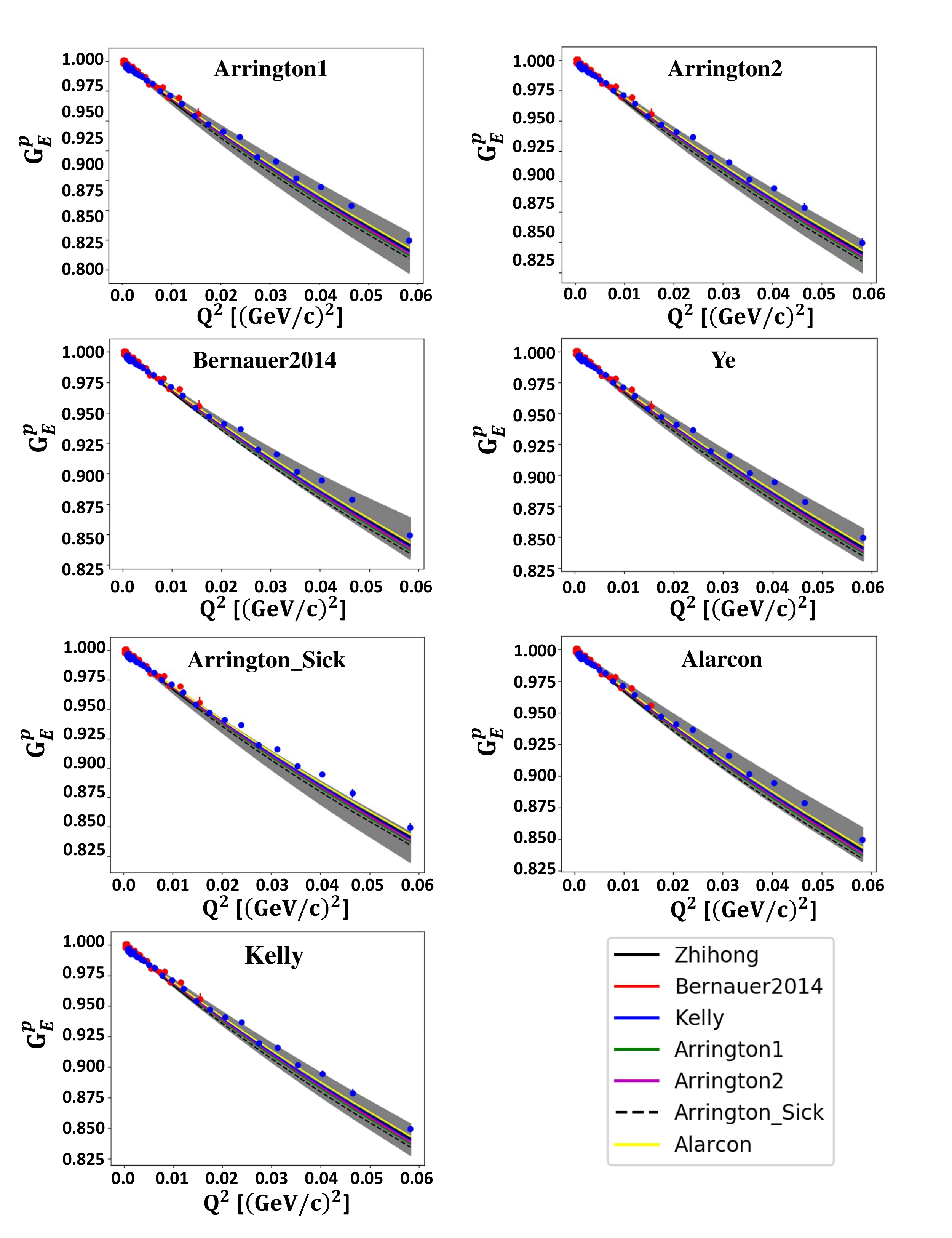}
\caption{(Color online) Seven proton electric form-factor models in which $G_{E}^{p}$ is plotted as a function of \(Q^2\). 
The gray bands are the bands generated by each smeared model. The superimposed red points are the PRad 1.1-GeV data; the 
blue points are the 2.2-GeV data \cite{Xiong:2019}. 
}
\label{fig:PRad_models}
\end{figure*}

The parameter smearing approach for deuteron form-factor models helps us better calculate the bias, by imitating a variety of reasonable approximations to the unknown true function, when the number of models is limited. In order to verify that this approach is valid and 
applicable, several proton electric form-factor \(G_{E}^{p}\) models can be tested in turn. Namely, we consider such parametrization 
models, including Kelly \cite{kelly2004simple}, Arrington1 \cite{Venkat:2010by}, Arrington2
\cite{Arrington:2003qk}, Arrington-Sick \cite{Arrington:2006hm}, Ye \cite{Ye:2017gyb}, Alarcon, and 
Bernauer-2014. The Alarcon model is our refit based upon \cite{Alarcon:2017ivh,Alarcon:2017lhg,Alarcon:2018irp}, 
and the Bernauer-2014 model is our refit of data from \cite{bernauer2014electric}. By smearing the parameters in the 
proton $G_{E}^{p}$ models, we determine whether or not the smearing method can mimic the low-$Q^{2}$ extrapolation behavior of 
those models.

Following the same steps shown in the previous section, the bias values obtained from fitting the Rational\,(1,1) with pseudo-data 
generated by the \(G_{E}^{p}\) models, before and after smearing, have been found and are displayed in Table~\ref{table2}. 
The nonsmeared bias in the table is the relative bias obtained by fitting pseudo-data generated from the original models. The smeared 
bias is the relative bias obtained by fitting pseudo-data generated from the smeared models following the procedure in the previous 
section. 

\begin{table}[h!]
\centering
  \begin{tabularx}{0.475\textwidth}{X X c}
      \hline
      \hline
      Model & \!\!\!\!\!\!\!\!\!\!Nonsmeared bias (\%) & Smeared bias (\%) \\
      \hline
    Kelly & ~~~~0.002 & 0.0007 \\
    Arrington1 & ~~~~0.005 & 0.003 \\
    Arrington2 & ~~~~0.009 & 0.002 \\
    Arrington-Sick & ~~~~0.001 & 0.0007 \\
    Alarcon & ~~~~0.166 & 0.174 \\
    Ye & ~~~~0.476 & 0.081 \\
    Bernauer-2014 & ~~~~0.271 & 0.062 \\
      \hline
      \hline
  \end{tabularx}
      \caption{The relative bias obtained from fitting the Rational\,(1,1) with pseudo-data generated by nonsmeared and 
      smeared seven proton $G_{E}^{p}$ models.}
\label{table2}
\end{table}

In Fig.~\ref{fig:PRad_models} we show a band of each model by smearing all the parameters (again in each model) for 
$\pm 10\%$, and restricting the values of \(\chi^{2}\) with respect to their degrees of freedom based on available data. 
One can also see that all the models and the superimposed PRad data are covered by most of the bands except for the 
band from the Arrington-Sick model, which means that the smearing method generates the pseudo-data in a reasonable range.

By looking at Table~\ref{table2} we find that the smeared bias is smaller than the nonsmeared bias for most 
of the models. This result is expected as the bias calculated from the smearing method gives the most probable value in the 
1$\sigma$ range based on the data. By looking at both Table~\ref{table2} and Fig.~\ref{fig:PRad_models}, we conclude that 
although the smearing method used with limited models can not precisely reflect the behavior of other models, 
it can exhibit more comprehensively how a fitter controls the bias.

\section*{Conclusions and outlook}
\begin{figure*}[hbt!]
\centering
\includegraphics[scale=0.315]{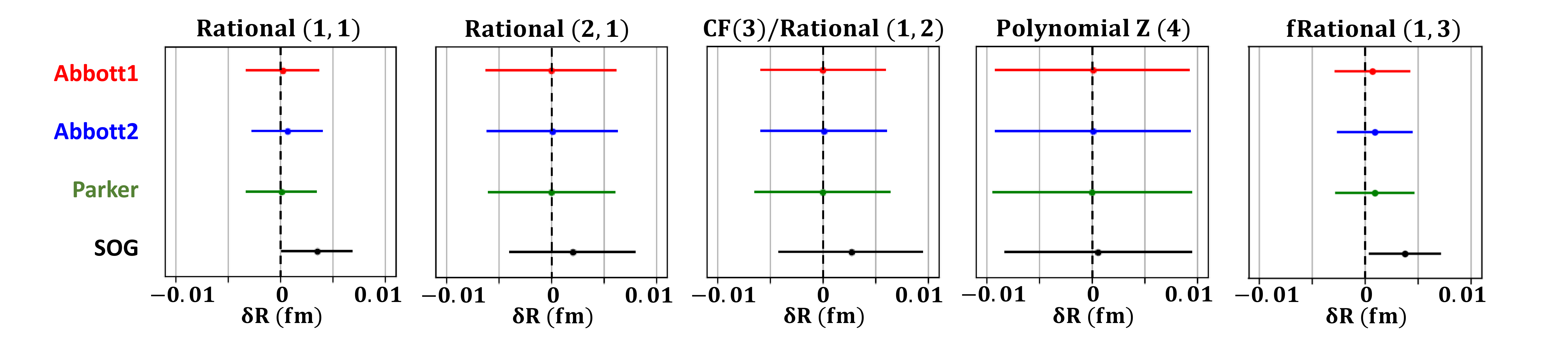}
\caption{(Color online) This figure shows the rms values of the bias for the shown fitters, derived from fitting pseudo-data 
generated by the four smeared Abbott1, Abbott2, Parker and SOG models (Sec.~\ref{sec:Smearing}). The error bars reflect the 
effects of the bin-by-bin total uncertainties of $G_C^d$ (Sec.~\ref{sec:procedure}).
}%
\label{fig:DRad_fitter_smeared1}%
\end{figure*}

In this section we summarize and conclude on our findings exhibited in the paper (including both appendices B and C). Also, we briefly discuss 
the prospects that this paper may have in the future.

Fig.~\ref{fig:DRad_fitter_smeared1} shows the rms values of the bias for the given five fitters, derived from fitting 
pseudo-data generated by the four smeared Abbott1, Abbott2, Parker and SOG models (see Sec.~\ref{sec:Smearing}), along with the bin-by-bin total 
uncertainties (see Sec.~\ref{sec:procedure}). According to the definition of the robustness discussed in Sec.~\ref{sec:robustness2}, the five 
fitters are all robust ($\rm bias[\rm rms] < \sigma_{stat}$). Although the Rational\,(1,1) and fRational\,(1,3) have larger 
bias values compared to those of the other three fitters, they can control the RMSE better because their variances are smaller 
than those of the others.

By comparing the bias and variance ($\sigma_{total}$) in that figure, our understanding is that the RMSE (overall
uncertainty) in the DRad experiment will be dominated by the bin-by-bin uncertainties rather than by the bias obtained in the fitting 
procedure. Based on our results, we propose to use the fRational\,(1,3) as the primary fitter in the deuteron charge radius extraction 
for this planned experiment, noting that it also has a better asymptotic behavior compared to that of Rational\,(1,1). Nonetheless, 
the fRational\,(1,3) is determined based on the data-driven method. Since it only has constraints from deuteron charge form-factor data at 
high $Q^{2}$, its extrapolation may not be very accurate, when it is used for fitting generated pseudo-data in a lower-$Q^{2}$ range. 
Once we have more data at low $Q^2$, we can better determine the fixed parameters in this fitter, in which case we will be able to 
extract the $r_{d}$ value more precisely. This might be done, for example, with possible upcoming new data from the A1 Collaboration at 
Mainz Microtron (MAMI). On the other hand, if we consider the results shown in Figs.~\ref{fig:R11_vs_FixedR13},~\ref{fig:DRad_fitter_smeared1},~\ref{fig:D1_R11_vs_modified},~and~\ref{fig:DRad_fitter_smeared2} together, 
in this case we find that (i) the fRational\,(1,3) and (ii) the modRational\,(1,1) are currently our best fitters for the robust extraction
of $r_{d}$. In addition, we note that the above-mentioned conclusions are anchored upon our studies for the DRad experiment. One should 
first account for the trade-off between the bias and variance, then select the best fitter stemming from the latest estimation of experimental 
uncertainties. If it turns out that the bin-by-bin uncertainties during the DRad experiment are much smaller (at least ten times) than 
what we have already evaluated, in this case we may search for another potentially robust fitter, which can minimize the bias and 
simultaneously will also have good asymptotics.

The radius extraction methods discussed so far depend on specific functional forms. In \cite{Craig2020fresh}, different extraction 
of the charge radius of the proton is discussed. The so-called cubic spline method is used to interpolate form-factor data, by which a 
smooth function is obtained afterwards. Then the radius could be extracted with an extrapolation using that smooth function. This method 
may also be applicable by us for the robust extraction of the deuteron charge radius in the near future, as an independent way for cross
checking our results coming from the ansatz provided in this paper.

\section*{Acknowledgments}
This work is supported in part by the U.S. Department of Energy under Grants No. DE-FG02-03ER41231 and 
No. DE-AC05-06OR23177, under which the Jefferson Science Associates operates the Thomas Jefferson National 
Accelerator Facility. This work is also supported in part by the U.S. National Science Foundation.

\appendix
\renewcommand{\theequation}{A\arabic{equation}}
\setcounter{equation}{0}
\section*{Appendix A: Details on the Abbott1, Abbott2, Parker, and SOG models}\label{sec:appA}

In this appendix we concisely discuss the parametrizations describing the Abbott1, Abbott2, Parker, and SOG models. \\

\subsection{Parametrization I~(Abbott1 model) \cite{Abbott:2000ak}} 
In the first parametrization, the charge form factor is represented by
\bea\label{D1}
& & G_{C}^{d}(Q^{2}) = 
\nonumber \\
& &
= G_{C,0}\times\left[1 - \left( \frac{Q}{Q^{0}_{C}} \right)^{2}\right]\times\left[1 + \sum_{i=1}^{5} a_{Ci}\,Q^{2i}\right]^{-1} ,
\eea
where \(G_{C,0}\) is a normalizing factor fixed by the deuteron charge, and \(Q_{C}^{0}\) and \(a_{Ci}\) are all together six 
free parameters that can be found on the website from \cite{Abbott:2000ak}. \\

\subsection{Parametrization II~(Abbott2 model) \cite{Abbott:2000ak,kobushkin1995deuteron}} 
The second parametrization is given by
\bea\label{D2}
& & G_{C}^{d}(Q^{2}) = 
\nonumber \\
& &~~~~~~~
= \frac{G^{2}(Q^{2})}{(2\tau + 1)}\left[\left( 1 - \frac{2}{3}\tau \right){g_{00}^+} + \frac{8}{3}\sqrt{2\tau}\,{g_{+0}^+} + 
\right.
\nonumber \\
& &~~~~~~~
\left. + \frac{2}{3}\left( 2\tau - 1 \right){g_{+ -}^+}\right] ,
\eea
where
\bea
& & {g_{00}^{+}} = \sum_{i=1}^{n}\frac{a_i}{\alpha^{2}_{i} + Q^{2}} , \quad {g_{+0}^{+}} = Q \sum_{i=1}^{n}\frac{b_i}
{\beta^{2}_{i}+ Q^{2}} , 
\nonumber \\
& &~~~~~~~~~~
\quad {g_{+-}^{+}} = Q^{2} \sum_{i=1}^{n}\frac{c_i}{\gamma^2_{i} + Q^{2}} .
\eea
$G(Q^2)$ in Eq.~(\ref{D2}) is a dipole form factor given by
\begin{equation}
G(Q^2) = \left( 1 + \frac{Q^2}{\delta^2} \right)^{-2} ,
\end{equation}
where $\delta$ is a parameter of the order of the nucleon mass.

The 24 parameters \({a_i},{b_i},{c_i},\alpha^{2}_{i}, \beta^{2}_{i} , \gamma^{2}_{i}\) can also be found on the website 
of \cite{Abbott:2000ak}. They are constrained by the following 12 relations:
\bea
& &\sum_{i=1}^{n}\frac{a_i}{\alpha_i^{2}} = 1 ,~~\sum_{i=1}^{n} b_{i}  = 0 ,~~~~
\quad \sum_{i=1}^{n}\frac{b_i}{\beta_i^{2}} = \frac{2 - \mu_{M}^{d}}{2\sqrt{2}M_{d}} ,
\nonumber \\
& &\sum_{i=1}^{n} c_{i} = 0 ,~\quad \sum_{i=1}^{n} c_{i}\gamma^{2}_{i} = 0 ,~
\quad \sum_{i=1}^{n}\frac{c_{i}}{\gamma_{i}^{2}} = \frac{1 - \mu_{M}^{d} - \mu_{Q}^{d}}{4M_{d}^2} ,
\nonumber \\
& &\alpha^{2}_{n} = 2M_{d}\,\mu^{(\alpha)} ,~\quad \alpha^{2}_{i} = 
\alpha^{2}_{1} + \frac{\alpha^{2}_{n} - \alpha^{2}_{1}}{n-1}(i-1) ,
\nonumber \\
& & \mbox{for}~\quad i = 1,...,n ,
\eea
where the parameter $\mu^{(\alpha)}$ has the dimension of energy and is of the order of $\Lambda_{QCD} \sim 0.2$~MeV. In total, there 
are 12 free parameters in this model. \\

\subsection{Parametrization III~(Parker model) \cite{Parker:2020}} 
The third parametrization is essentially based upon the remade fits from the first two parametrizations, however, with
constraints to prevent singularities in the functional forms of the $G_{C}^{d}$, $G_{Q}^{d}$ and $G_{M}^{d}$ form factors:
\bea\label{Parker}
& & G_{C}^{d}(Q^{2}) = 
\nonumber \\
& &
G_{C,0}\times\left[1 - \left( \frac{Q}{Q^{0}_{C}} \right)^{2}\right]\times\left[\prod_{i=1}^{5} (1 + |a_{i}|\,Q^2) \right]^{-1} ,
\eea
where the values of \(G_{C,0}\) and \(Q_{C}^{0}\) are the same as the ones shown in Eq.~(\ref{D1}). \(a_{i}\) are all together five free parameters 
determined from fitting the data from the website of \cite{Abbott:2000ak}. \\

\subsection{Parametrization IV~(SOG model) \cite{Abbott:2000ak,Sick:1974suq,Zhou:2020}} 
The fourth parametrization utilizes the SOG method, by which $G_{C}^{d}(Q^{2})$ reads as
\bea\label{SOG}
& & G_{C}^{d}(Q^{2}) = 
\nonumber \\
& &
= G_{C,0}\times e^{-\frac{1}{4}Q^2 \gamma^2}\times \sum_{i=1}^{N} \frac{A_i}{1 + 2R_{i}^{2}/\gamma^{2}} \times
\nonumber \\
& &~~~~~~~
\times \left[\cos{\!(QR_{i})} + \frac{2R_{i}^{2}}{\gamma^2}\frac{\sin{\!(QR_{i})}}{QR_{i}}\right] .
\eea
In the configuration space this parametrization corresponds\footnote{The density $\rho$ is a function of the distance $s$, 
which is the distance of the nucleons to the deuteron center of mass.} to a density $\rho(s)$ given in terms of a sum of Gaussians located at arbitrary radii $R_{i}$, 
with amplitudes $A_{i}$ fitted to the data and with a fixed width $\gamma$, where $\gamma\sqrt{3/2} = 0.8$~fm.

In our fitting we take $N = 12$. There are 11 free fitting parameters: ten Gaussian amplitudes $A_1$, $A_2$, ..., $A_{10}$ that correspond to ten
$R_{i} < 4{~\rm fm}$, and one overall amplitude $A_{11}$ corresponding to the range of $R_{11}$ from $4$ to $10~{\rm fm}$. For obtaining 
the normalization, there is one more amplitude $A_{12}$ with $R_{12} = 0.4~{\rm fm}$. All the amplitudes satisfy the condition 
$\sum_{i=1}^{12} A_{i} = 1$. To determine the parameters $A_i$, a set of $R_i$ is randomly generated in the range mentioned above, then 
the function in Eq.~(\ref{SOG}) is fitted to the $G_{C}^{d}$ data set from Table~1 of \cite{Abbott:2000ak}. The sets of 
$R_i$ are generated repeatedly until the $\chi^2$ value is minimized and converged. With 11 fixed $R_i$ and 11 free 
parameters $A_i$, a fit to the data set is obtained with $\chi^2/{\rm NDF} \simeq 1.63$ \cite{Zhou:2020}.

\appendix
\renewcommand{\theequation}{B\arabic{equation}}
\setcounter{equation}{0}
\section*{Appendix B: A fitter test based upon using theory-based models}\label{sec:appB}

Except for the four data-based deuteron charge form-factor models under consideration, we also test some theory-based models 
by following the same method developed in \cite{Yan:2018bez}. The data generation and fitting procedure is already 
described in Sec.~\ref{sec:procedure} with statistical fluctuations included. Here we use the following $G^{d}_{C}$ models as 
additional generators.

(i) The IA (relativistic impulse approximation), IAMEC (relativistic IA plus meson exchange current), RSC (relativistic 
IA with Reid Soft Core), and RSCMEC (relativistic IA plus meson exchange current with Reid Soft Core) are the parametrizations to the theoretical calculations discussed in \cite{Hummel:1993fq}.

(ii) The quadratic and cubic models are the second- and third-order polynomial fits to theoretical points calculated by using 
the model in \cite{Gross:2019thk}, where the parameters in these models are given in Table~13 of \cite{Gross:2019thk}. 

(iii) The Gaussian, Monopole and Dipole are naive models that imitate possible approximations to the would-be true 
form-factor function (from Nature) at $Q^{2} \rightarrow 0$. Their 
functional forms can be found in \cite{Yan:2018bez}. 

\begin{figure}[hbt!]
\centering
\includegraphics[width=0.47\textwidth]{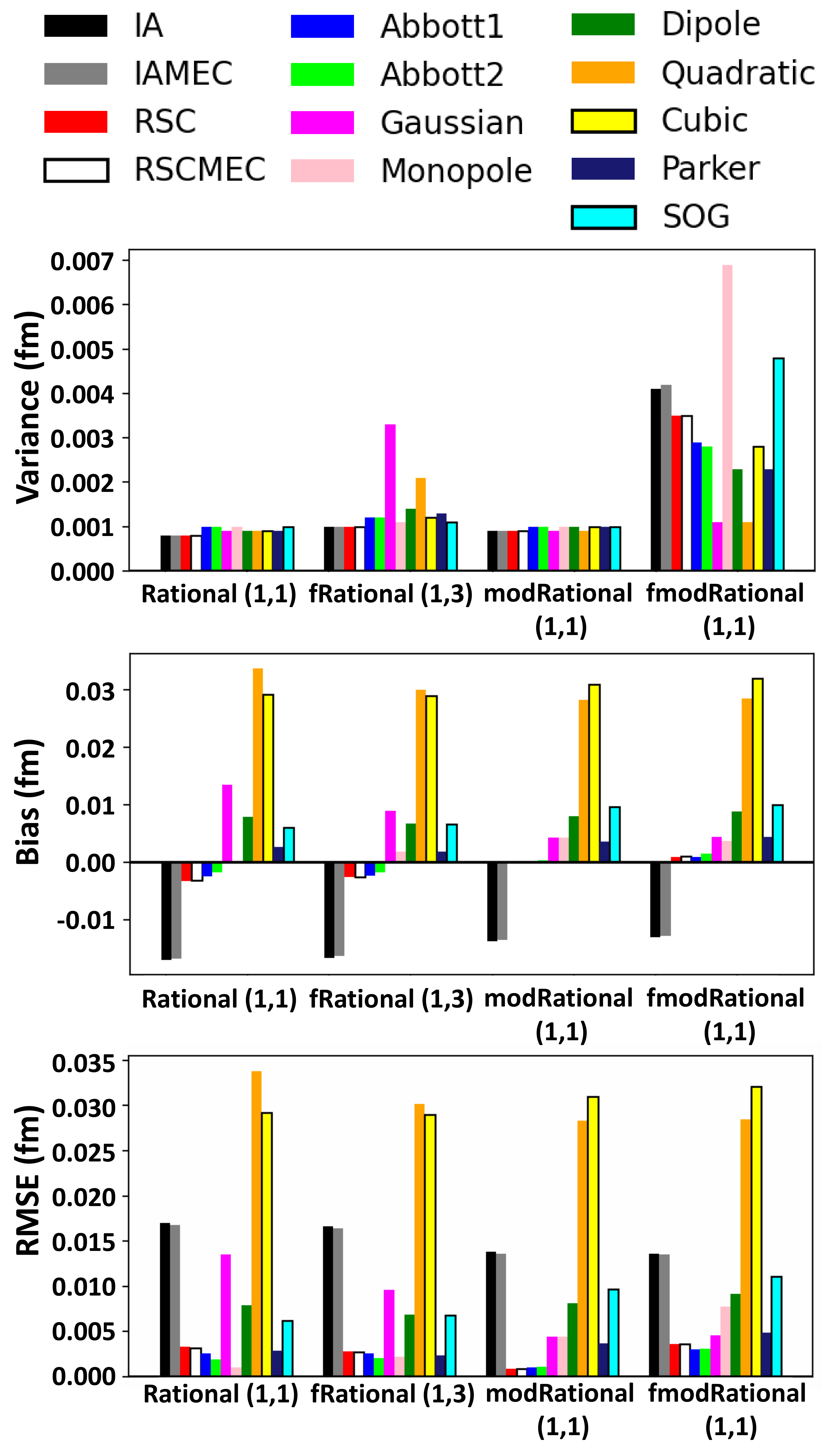}
\caption{(Color online) The variance (rms value) and bias obtained from fitting the given fitters with pseudo-data (including 
statistical fluctuations) generated by 11 different deuteron models. The RSCMEC, RSC, IAMEC, and IA models are discussed in 
\cite{Hummel:1993fq}. Both Abbott parametrizations are taken from \cite{Abbott:2000ak,kobushkin1995deuteron}; the
Parker and SOG parametrizations are taken from \cite{Parker:2020} and \cite{Sick:1974suq}, respectively. The Dipole, Monopole,
and Gaussian are described by simple models. The Quadratic/Cubic models are taken from \cite{Gross:2019thk}. The RMSE calculation 
here follows \cite{Yan:2018bez}.
}
\label{fig:fit_models}
\end{figure}

Fig.~\ref{fig:fit_models} shows the statistical variance, bias, and their quadratic sum (RMSE) from fitting various fitters with 
pseudo-data (including statistical fluctuations) generated by 11 different deuteron models. One can see that the bias 
is much larger than the statistical variance. 
In Table~\ref{tab:table_chi2} we show the calculated $\chi^2$ values [from Eq.~(\ref{Eq:chi2})] for all the models from 
Fig.~\ref{fig:fit_models} using the available data points from Table~1 of \cite{Abbott:2000ak}, and find that those 
 theory-based models have large discrepancies with the experimental data. We conclude that the method to test the 
robustness in the charge radius extraction of the proton from \cite{Yan:2018bez} is not suitable for the deuteron's case in the 
DRad kinematics. When we investigate the properties of the fitters for DRad, we would need to have more data to decide whether 
we should take any of these existing theory-based models into consideration. 

\begin{table}[hbt!]
\centering
  \begin{tabularx}{0.35\textwidth}{X c}
    \hline
    \hline
      Models ~&~$\chi^2$ \\
      \hline
    IA       &183.09 \\
    IAMEC    &295.57 \\
    RSC      &125.74 \\
    RSCMEC   &196.71 \\
    Abbott1  &19.77 \\
    Abbott2  &38.59 \\
    Parker  &22.94 \\
    SOG  &17.88 \\
    Gaussian &668.59 \\
    Monopole &$1.14 \times 10^{5}$ \\
    Dipole   &$5.83 \times 10^{3}$ \\
    Quadratic &$2.65 \times 10^{12}$ \\
    Cubic &$1.15 \times 10^{15}$ \\
    \hline
    \hline
  \end{tabularx}
      \caption{The $\chi^2$ value, obtained from Eq.~(\ref{Eq:chi2}), for each of the 11 models using the available data points 
      from Table~1 of \cite{Abbott:2000ak}. }
\label{tab:table_chi2}
\end{table}

\appendix
\renewcommand{\theequation}{C\arabic{equation}}
\setcounter{equation}{0}
\section*{Appendix C: Searching for other robust fitter candidates}\label{sec:appC}

\subsection{The modified Rational\,(1,1) function}

Except for the fRational\,(1,3) fitter function discussed in the paper, for the deuteron charge radius extraction we have also studied 
a modified and generalized version of Rational\,(1,1), which we designate as modified Rational\,(1,1) [or simply as modRational\,(1,1)]:
\bea\label{modified_R11}
& & f_{\rm modified\,Rational\,(1,1)}(Q^{2}) \equiv {\rm modRational\,(1,1)} = 
\nonumber \\
& &
~~~~~~~~~~~~ = p_{02}\,G_{C}^{d}(Q^{2}) =
p_{02}\frac{\left( 1 + p_{1}^{(a^{\prime})}Q^{2} \right)^{A}}
{\left( 1 + p_{1}^{(b^{\prime})}Q^{2} \right)^{B}} ,
\eea
where \(p_{02}\) is a floating normalization parameter, and $p_{1}^{(a^{\prime})}$ and $p_{1}^{(b^{\prime})}$ are two free fitting parameters. 
To control the variance, we need to limit the number of the free parameters. The deuteron rms charge radius is calculated by 
\(r_{d} = \sqrt{6 \left( B \times {p_{1}^{(b^{\prime})}} - A \times {p_{1}^{(a^{\prime})}} \right)}\). It is obvious that 
Eq.~(\ref{modified_R11}) reduces to Eq.~(\ref{R11}) at $A$ = 1 and $B$ = 1. Both powers $A$ and $B$ can be fixed and given by different methods 
that we discuss below.

To search for the best combination of $A$ and $B$ in Eq.~(\ref{modified_R11}), for the purpose of extracting $r_{d}$ robustly within 
the scope of the data-based models, we use a scanning approach. In this approach, $A$ and $B$ (each) are varied from 0 to 10 with 
the step equal to 0.1, and the fitter is fitted with the pseudo-data generated by the four models discussed in Appendix A in the DRad kinematic range of 
$2 \times 10^{-4}~{\rm (GeV/c)^2} < Q^{2} < 0.05~{\rm (GeV/c)^2}$. Using the scanning approach, we obtain $A = 3.0 - 4.2$ and $B = 0.8$ 
to be the best $A$ range and $B$ value to minimize the bias.

Nonetheless, the outlined scanning method is model-dependent, which is limited by the number of the given charge form-factor models. 
In that case the fewer the reliable models are, the higher the model-dependency is. In order to avoid this issue, we have also tried 
a data-driven method described in Sec.~\ref{sec:Datadriven}. For the modRational\,(1,1), when $A$ and $B$ are also considered as free
parameters, there are in total four free parameters. We use this fitter for fitting the form-factor data at the high-$Q^{2}$ region listed 
in Table 1 of \cite{Abbott:2000ak} -- which gives {$A$ = \(3.48668 \pm 0.01568 \) and $B$ = \(0.75600 \pm 0.11313 \)} -- then fix 
these values for fitting the pseudo-data in the low-$Q^{2}$ range of the DRad kinematics. In this case we will have  the fixed modified 
Rational\,(1,1) [or simply the fmodRational\,(1,1)]:
\bea\label{modified3_R11}
& & f_{\rm fixed\,modified\,Rational\,(1,1)}(Q^{2}) \equiv {\rm fmodRational\,(1,1)} =  
\nonumber \\
& & 
~~~~~~~~~~~~~~~~~ = p_{02}\frac{\left( 1 + p_{1}^{(a^{\prime})}Q^{2} \right)^{A,{\rm fixed}}}
{\left( 1 + p_{1}^{(b^{\prime})}Q^{2} \right)^{B,{\rm fixed}}} ,
\eea
where the uncertainties in the fixed parameters are taken also into account when we calculate the bias. 

\begin{figure}[hbt]
\centering
\includegraphics[width=0.45\textwidth]{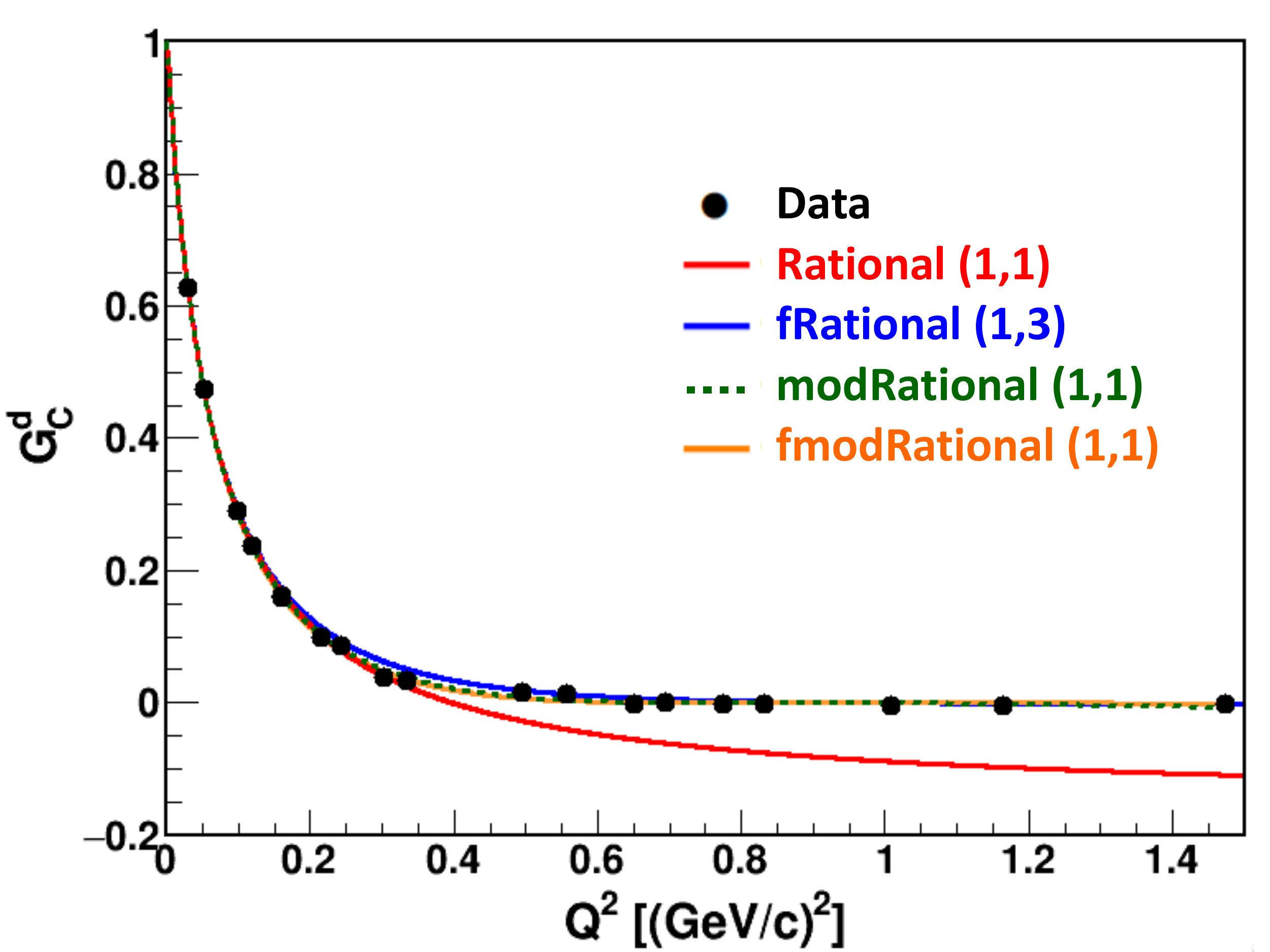}
\includegraphics[width=0.48\textwidth]{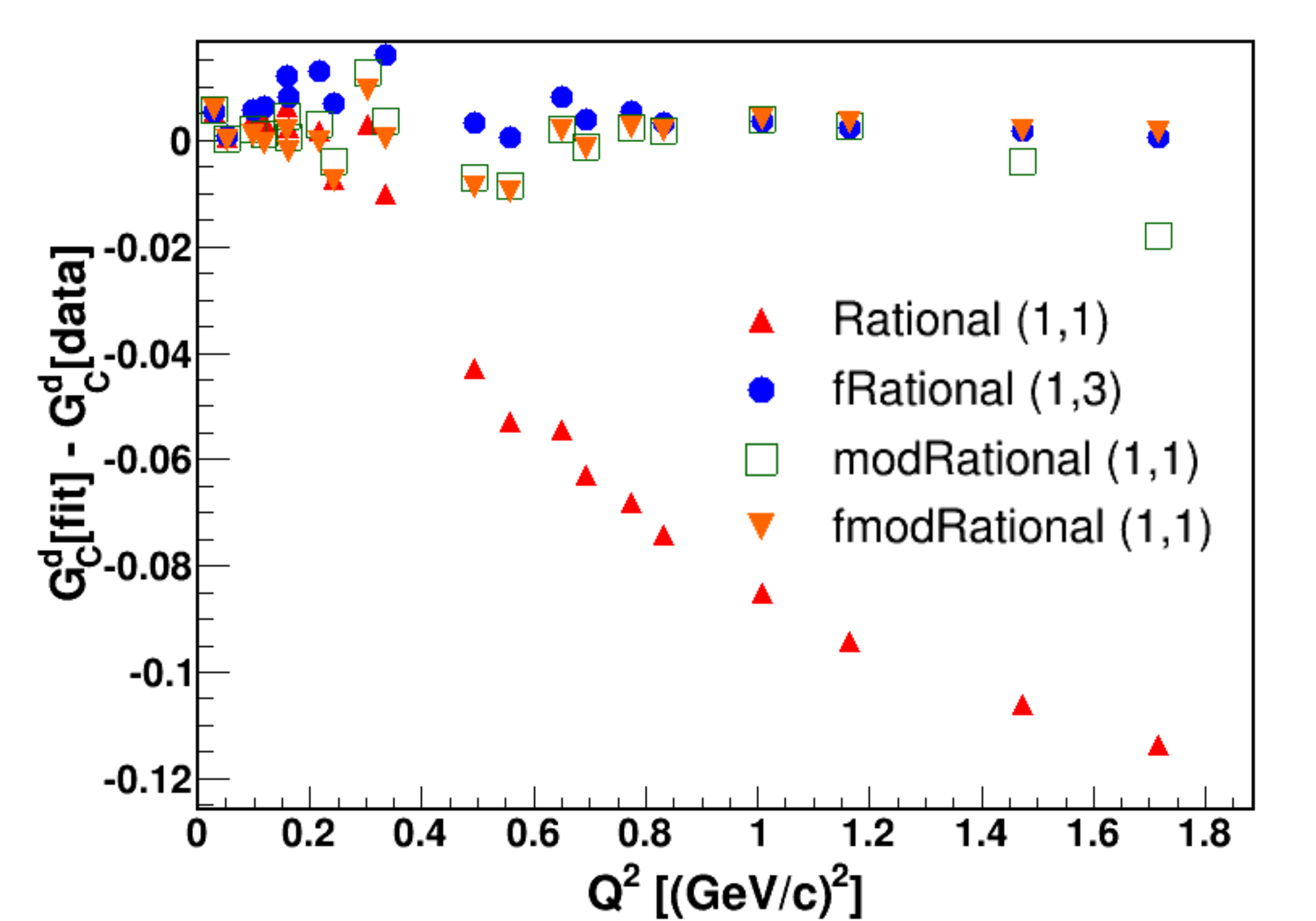}
\caption{(Color online) 
The upper plot shows the Rational\,(1,1), 
fRational\,(1,3), modRational\,(1,1) (with $A$ = 3.4 and $B$ = 0.8), and fmodRational\,(1,1) (with $A$ =  \(3.48668 \) 
and $B$ = \(0.75600 \)) obtained from fitting the pseudo-data generated by the Abbott1 model \cite{Abbott:2000ak},
which for comparison are overlaid with the black colored data points listed in Table 1 of \cite{Abbott:2000ak}. 
The color coding is displayed in the legends, where the CF\,(3) and Rational\,(1,2) are the same and described by the 
two asymptotic green dotted lines. The lower plot shows the residual points for these fitters, where ``the residual" 
means the difference between $G_{C}^{d}[{\rm fit}]$ described by the fitters and $G_{C}^{d}[{\rm data}]$ from the data.} 

\label{fig:D1_R11_vs_modified}
\end{figure}

To compare the differences between the Rational\,(1,1), fRational\,(1,3), modRational\,(1,1), and fmodRational\,(1,1) in the Abbott1/Abbott2 
model range [from \(3 \times 10^{-2}~{\rm to}\ 1.5~{\rm (GeV/c)^2}\)], all the functions are plotted in this range. As an
example, we pick up fixed values $A = 3.4$ and $B = 0.8$ in the modRational\,(1,1) in Eq.~(\ref{modified_R11}). The parameters in these different 
fitters are determined by fitting pseudo-data generated from the Abbott1 model in the DRad range. The results from the Abbott2, 
Parker, and SOG models are quite similar, and are not shown here. As shown in Fig.~\ref{fig:D1_R11_vs_modified}, except for the Rational\,(1,1), 
the other fitters show good asymptotic behavior in the high-\(Q^{2}\) range.

\subsection{Similarity of the fitters modRational\,(1,1) and Rational\,(1,3)}

The modRational\,(1,1) fitter lacks a clear physical meaning. Meanwhile, one can show that the functional form of the 
modRational\,(1,1) is actually similar to the Ratioanl\,(1,3). This could be demonstrated if we started with the fitted modRational\,(1,1) 
for generating a set of $G_{C}^{d}$ pseudo-data, and then used the Rational\,(1,3) for fitting this set of generated pseudo-data.
Thereby, we have the following steps.

(i) the modRational\,(1,1) with $A$ = 3.4 and $B$ = 0.8 from Eq.~(\ref{modified_R11}) is used to fit pseudo-data generated 
by the Abbott1 model \cite{Abbott:2000ak}. The fitted function comes out to be the following:
\begin{equation}\label{fitmodR11}
~~~~~~~~~~
{\rm modRational~(1,1)} = \frac{\left( 1 - 0.0456785\,Q^2 \right)^{3.4}}
{\left( 1 + 0.718695\,Q^{2} \right)^{0.8}},
\end{equation}
where the dimension of $Q^{2}$ is in ${\rm fm^{-2}}$.

(ii) to generate a set of $G_{C}^{d}$ pseudo-data with reasonable bins and uncertainties, we choose both the DRad binning 
with its simulated statistical uncertainty and the Abbott binning from Table~1 of \cite{Abbott:2000ak}. In total, there 
are 82 pseudo-data points that are generated by the function in Eq.~(\ref{fitmodR11}), in the range of 
$Q^{2} = 0.006 - 21.344$~${\rm fm}^{-2}$.

(iii) the Rational\,(1,3) function as shown in Eq.~(\ref{R13}) is used to fit those pseudo-data.

\begin{figure}[hbt!]
\centering
\includegraphics[scale=0.38]{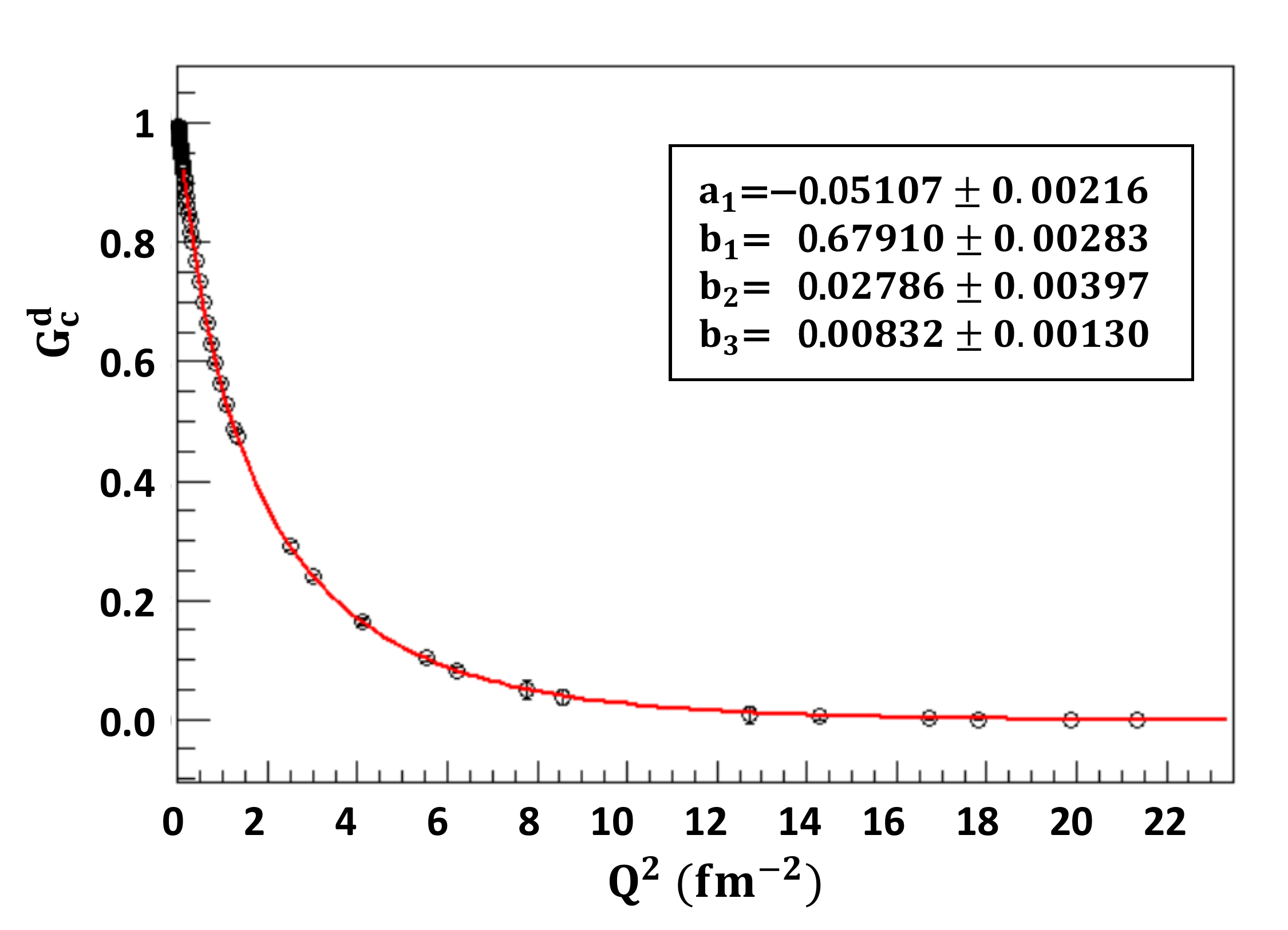}
\caption{(Color online) The Rational (1,3) (red curve) fitted with the pseudo-data generated by Eq.~(\ref{fitmodR11}) (black points).}
\label{fig:ModR11_R13}
\end{figure}

By fulfilling the above steps, we present the result in Fig~\ref{fig:ModR11_R13}, where the black points are the pseudo-data 
points generated from Eq.~(\ref{fitmodR11}), and the red curve is a fitted Rational\,(1,3). This figure shows that the 
modRational\,(1,1) has a very similar behavior as the Rational\,(1,3) in the range of $Q^{2} < 21.5~{\rm fm}^{-2}$ or 
equivalently of $Q^{2} < 0.84~({\rm GeV/c})^{2}$.

\begin{figure*}[hbt!]
\centering
\includegraphics[scale=0.35]{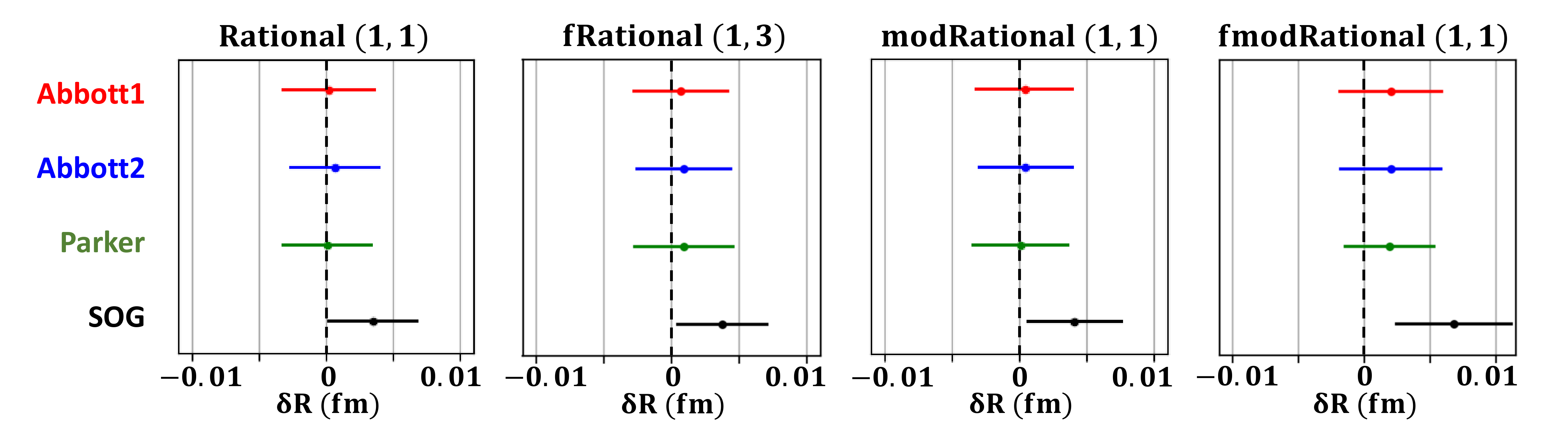}
\caption{(Color online) This figure shows the rms values of the bias for the shown fitters, derived from fitting pseudo-data 
generated by the four smeared Abbott1, Abbott2, Parker, and SOG models (Sec.~\ref{sec:Smearing}). The error bars reflect the 
effects of the bin-by-bin total uncertainties of $G_C^d$ (Sec.~\ref{sec:procedure}).
}%
\label{fig:DRad_fitter_smeared2}%
\end{figure*}

Finally, we wish to briefly discuss Fig.~\ref{fig:DRad_fitter_smeared2}, which shows the rms values of the bias for the given 
four fitters, derived from fitting pseudo-data generated by the four smeared Abbott1, Abbott2, Parker, and SOG models 
(see Sec.~\ref{sec:Smearing}), along with the bin-by-bin total uncertainties (see Sec.~\ref{sec:procedure}). For the modRational\,(1,1), 
the results of various tested combinations with $A =3.0 - 4.2$ and $B$ = 0.8 are stable inside the $A$ range. As shown in this figure, all 
the fitters are robust ($\rm bias[\rm rms] < \sigma_{stat}$) based on the definition in Sec.~\ref{sec:robustness2}. The RMSE (overall uncertainty) 
values from the Rational\,(1,1), fRational\,(1,3), and modRational\,(1,1) are similar, which means that the modRational\,(1,1) can also be 
considered as a good fitter candidate for the deuteron charge radius extraction. However, taking also into account the $G_{C}^{d}$ behavior that 
we observe in Fig.~\ref{fig:R11_vs_FixedR13} and Fig.~\ref{fig:D1_R11_vs_modified}, we consider both fRational\,(1,3) and modRational\,(1,1) as 
the best two fitters for the robust extraction of $r_{d}$ in the DRad experiment because the Rational\,(1,1) drops out in this combined picture.


\begin{thebibliography}{99}    

\bibitem{Miller:2018ybm}
G.~A.~Miller,
Phys. Rev. C \textbf{99}, no.3, 035202 (2019).

\bibitem{Pohl:2010} 
R.~Pohl {\it et al.}, Nature (London) {\bf 466}, 213 (2010).

\bibitem{Antognini:1900n} 
A.~Antognini {\it et al.}, Science {\bf 339}, 417 (2013).

\bibitem{Mohr:2015ccw} 
P.~J.~Mohr, D.~B.~Newell and B.~N.~Taylor,
Rev.\ Mod.\ Phys.\ {\bf 88}, 035009 (2016).

\bibitem{Pohl:2013yb} 
R.~Pohl, R.~Gilman, G.~A.~Miller and K.~Pachucki,
Ann.\ Rev.\ Nucl.\ Part.\ Sci.\  {\bf 63}, 175 (2013).

\bibitem{Carlson:2015jba} 
C.~E.~Carlson,
Prog.\ Part.\ Nucl.\ Phys.\ {\bf 82}, 59 (2015).

\bibitem{Hill:2017wzi} 
R.~J.~Hill,
EPJ Web Conf.\  {\bf 137}, 01023 (2017).

\bibitem{Fleurbaey:2018} 
H.~Fleurbaey, S.~Galtier, S.~Thomas, M.~Bonnaud, L.~Julien, F.~Biraben, F.~Nez, M.~Abgrall, and J.~Guena, Phys.\ Rev.\ Lett.\ {\bf 120}, 183001 (2018).

\bibitem{Beyer:2017} 
A.~Beyer {\it et al.}, Science {\bf 358}, 79 (2017).

\bibitem{Bezginov:2019} 
N. Bezginov, T. Valdez, M. Horbatsch, A. Marsman, A. C. Vutha, E. A. Hessels, Science {\bf 365}, 1007 (2019).

\bibitem{Grinin:2020} 
A. Grinin, A. Matveev, D. C. Yost, L. Maisenbacher, V. Wirthl, R. Pohl, T. W. H\"{a}nsch,
T. Udem, Science {\bf 370}, 1061 (2020).

\bibitem{Xiong:2019} 
W.~Xiong {\it et al.}, Nature (London) {\bf 575}, 147 (2019).

\bibitem{Gasparian:2014rna} 
A.~Gasparian (PRad at JLab), 
EPJ Web Conf. {\bf 73}, 07006 (2014).

\bibitem{Peng:2015szv} 
C.~Peng and H.~Gao, 
EPJ Web Conf. {\bf 113}, 03007 (2016).

\bibitem{Pohl:2016} 
R.~Pohl {\it et al.}, Science {\bf 353}, 669 (2016).

\bibitem{Mohr:2012} 
P.~J.~Mohr, B.~N.~Taylor and D.~B.~Newell,
Rev.\ Mod.\ Phys.\ {\bf 84}, 1527 (2012).

\bibitem{Pohl:2016glp} 
R.~Pohl {\it et al.},
Metrologia {\bf 54}, L1 (2017).

\bibitem{Parthey:2010aya} 
C.~G.~Parthey, A.~Matveev, J.~Alnis, R.~Pohl, T.~Udem, U.~D.~Jentschura, N.~Kolachevsky and 
T.~W.~H\"{a}nsch
Phys.\ Rev.\ Lett.\ {\bf 104}, 233001 (2010).

\bibitem{DRad} 
PRad Collaboration, Precision Deuteron Charge Radius Measurement with Elastic Electron-Deuteron Scattering, 
\url{https://www.jlab.org/exp_prog/proposals/17/PR12-17-009.pdf}.

  
\bibitem{Abbott:2000ak} 
D.~Abbott {\it et al.} (JLab t20 Collaboration),
Eur.\ Phys.\ J.\ A {\bf 7}, 421 (2000).

\bibitem{Abbott:2000fg} 
D.~Abbott {\it et al.} (JLab t20 Collaboration),
Phys.\ Rev.\ Lett.\  {\bf 84}, 5053 (2000).

\bibitem{kobushkin1995deuteron} 
A.~P.~Kobushkin and A.~I.~Syamtomov, Phys.\ Atom.\ Nucl.\ {\bf 58}, 1477 (1995) [Yad.\ Fiz.\ {\bf 58N9}, 1565 (1995)].

\bibitem{Parker:2020} 
A.~Parker and D.~W.~Higinbotham,  Deuteron Form Factor Parameterization (2020),
\url{https://doi.org/10.5281/zenodo.4074280}.

\bibitem{Sick:1974suq}
I.~Sick,
Nucl. Phys. A \textbf{218}, 509 (1974).

\bibitem{Zhou:2020} 
J.~Zhou, The Sum-of-Gaussian parameterizations fitted with the available deuteron form factor data (2020),
\url{https://github.com/TooLate0800/Deuteron_radius_fitting/tree/master/SOG_fitting}.

\bibitem{Yan:2018bez} 
X.~Yan {\it et al.},
Phys.\ Rev.\ C {\bf 98}, 025204 (2018).

\bibitem{Wong:1994sy} 
C.~W.~Wong,
Int.\ J.\ Mod.\ Phys.\ E {\bf 3}, 821 (1994).

\bibitem{Berard:1974ev} 
R.~W.~Berard, F.~R.~Buskirk, E.~B.~Dally, J.~N.~Dyer, X.~K.~Maruyama, R.~L.~Topping and T.~J.~Traverso,
Phys.\ Lett.\ 47 {\bf B}, 355 (1973).

\bibitem{Simon:1981br} 
G.~G.~Simon, C.~Schmitt and V.~H.~Walther,
Nucl.\ Phys.\ A {\bf 364}, 285 (1981).

\bibitem{Platchkov:1989ch} 
S.~Platchkov {\it et al.},
Nucl.\ Phys.\ A {\bf 510}, 740 (1990).

\bibitem{Jankus:1997} 
V.~Z.~Jankus, Phys.\ Rev.\ {\bf 102}, 1586 (1956).

\bibitem{Gourdin:1963} 
M.~Gourdin, Nuov.\ Cim.\ {\bf 28}, 533 (1963).

\bibitem{Mainz} 
Mainz Microtron [MAMI], Measurement of the elastic $A(Q^{2})$ form factor of the deuteron 
at very low momentum transfer and the extraction of the monopole charge radius of the deuteron, 
\url{http://wwwa1.kph.uni-mainz.de/A1/publications/proposals/MAMI-A1-01-2012.pdf}.

\bibitem{Garcon:2001sz} 
M.~Garcon and J.~W.~Van Orden,
Adv.\ Nucl.\ Phys.\  {\bf 26}, 293 (2001).

\bibitem{Kraus:2014qua}
E.~Kraus, K.~E.~Mesick, A.~White, R.~Gilman and S.~Strauch,
Phys. Rev. C \textbf{90}, 045206 (2014).

\bibitem{Radius_fitting_lib} 
Proton radius fitting library, \url{https://github.com/saberbud/Proton_radius_fit_class}.

\bibitem{Brun:1997} 
R.~Brun and F.~Rademakers, Nucl.\ Instrum.\ Meth.\ A {\bf 389}, 81 (1997).

\bibitem{James:1975} 
F.~James and M.~Roos, Comput.\ Phys.\ Commun.\ {\bf 10}, 343 (1975).

\bibitem{Higinbotham:2019jzd}
S.~K.~Barcus, D.~W.~Higinbotham and R.~E.~McClellan,
Phys. Rev. C \textbf{102}, 015205 (2020).

\bibitem{HTF:2009}
T.~Hastie, R.~Tibshirani and J.~Friedman,
The elements of statistical learning: data mining, inference and prediction, 2nd ed., (Springer, New York, 2009).

\bibitem{Lee:2015jqa} 
G.~Lee, J.~R.~Arrington and R.~J.~Hill,
Phys.\ Rev.\ D {\bf 92}, 013013 (2015).

\bibitem{Binningset} 
DRad bin-set files, 
\url{https://github.com/TooLate0800/Deuteron_radius_fitting}.

\bibitem{kelly2004simple} 
J.~J.~Kelly,
Phys. Rev. C \textbf{70}, 068202 (2004).

\bibitem{Venkat:2010by}
S.~Venkat, J.~Arrington, G.~A.~Miller and X.~Zhan,
Phys. Rev. C \textbf{83}, 015203 (2011).

\bibitem{Arrington:2003qk}
J.~Arrington,
Phys. Rev. C \textbf{69}, 022201(R) (2004).

\bibitem{Arrington:2006hm}
J.~Arrington and I.~Sick,
Phys. Rev. C \textbf{76}, 035201 (2007).

\bibitem{Ye:2017gyb}
Z.~Ye, J.~Arrington, R.~J.~Hill and G.~Lee,
Phys. Lett. B \textbf{777}, 8 (2018).

\bibitem{Alarcon:2017ivh}
J.~M.~Alarc\'{o}n and C.~Weiss,
Phys. Rev. C \textbf{96}, 055206 (2017).

\bibitem{Alarcon:2017lhg}
J.~M.~Alarc\'{o}n and C.~Weiss,
Phys. Rev. C \textbf{97}, 055203 (2018).

\bibitem{Alarcon:2018irp}
J.~M.~Alarc\'{o}n and C.~Weiss,
Phys. Lett. B \textbf{784}, 373 (2018).

\bibitem{bernauer2014electric} 
J.~C.~Bernauer, M.~O.~Distler, J.~Friedrich, T.~Walcher, P. ~Achenbach, C.~Ayerbe Gayoso {\it et al.} (A1 Collaboration), Phys.\ Rev.\ C {\bf 90}, 015206 (2014).

\bibitem{Craig2020fresh} 
C.~D.~Roberts, Private communication.

\bibitem{Hummel:1993fq}
E.~Hummel and J.~A.~Tjon,
Phys. Rev. C \textbf{49}, 21 (1994).

\bibitem{Gross:2019thk}
F.~Gross,
Phys. Rev. C \textbf{101}, 024001 (2020).

\end{thebibliography}
\end{document}